\documentclass[final,3p,times]{elsarticle}
\usepackage{amsmath,amssymb,amsbsy,amsfonts,amsthm,mathrsfs}
\usepackage{bm,bbm,graphicx}
\usepackage{float}
\usepackage{epsfig}
\usepackage{color}
\usepackage[dvipsnames]{xcolor}
\usepackage[colorlinks=true,citecolor=Cerulean,linkcolor=RubineRed,urlcolor=Cerulean]{hyperref}
\usepackage{silence}
\journal{Physica A: Statistical Mechanics and its Applications}
\usepackage{soul}
\usepackage{xfrac}
\usepackage{subfigure}
\usepackage[normalem]{ulem}
\usepackage[version=4]{mhchem}
\usepackage{stackrel}
\usepackage{lineno,hyperref}
\modulolinenumbers[5]

\begin{document}
	
\begin{frontmatter}
\title{Magnetic field induced anomalous distribution of particles}

	\author[label1]{Shrabani~Mondal}
	\address[label1]{Department of Chemistry,\
		University of Massachusetts Boston,\
		Boston, MA 02125}
	
\author[label2]{L.~R.~Rahul~Biswas}
\address[label2]{Department of Chemistry,\
	Visva-Bharati, Santiniketan,\
	India, 731235}	
	
\author[label2]{Mousumi~Biswas}

	\author[label2]{Bidhan~Chandra~Bag}
	\cortext[mycorrespondingauthor]{Corresponding author}
	\ead{bidhanchandra.bag@visva-bharati.ac.in}

\begin{abstract}

It seems that at least the deterministic or the noise part of a stochastic system must be a nonlinear function of phase space variable(s) to observe the phenomenon, noise induced transition. But in the present paper, we have demonstrated that the phenomenon may be observed even in a linear stochastic process where both the deterministic and the stochastic parts are linear functions of the relevant phase space variables. The shape of the stationary distribution of particles (which are confined in a harmonic potential) may change on increasing the strength of the applied fluctuating magnetic field (FMF). The probability density may vary non monotonically with an increase in the coordinate of a Brownian particle. Thus the distribution of particles may deviate strongly from the Boltzmann one and it may be a unique signature of the FMF with a new mechanism in the field of noise induced transition. Then we are motivated strongly to study the distribution of particles in a nonlinear stochastic system where the Brownian particles are confined in a bi-stable potential energy field in the presence of the fluctuating magnetic field. With a relatively large strength of the fluctuating field, the distribution of particles may be such that where many islands may appear which are not expected from the given potential energy field. It may offer an explanation to describe the phenomenon, the reduction of the current in a semiconductor in the presence of a time dependent magnetic field.
\end{abstract}

\begin{keyword}
		Brownian particle,  Fluctuating magnetic field, noise induced transition, anomalous distribution
\end{keyword}

\end{frontmatter}


\section{Introduction}
It is commonly believed that noise is a phenomenon which induces disorder. But there are situations where it plays a constructive role such as stochastic resonance \cite{gamma,welle,raja,shm}, coherence resonance  \cite{pikov}, resonant activation \cite{gamma,doer,majee,hu,huI,sr,bag6}, Brownian ratchets \cite{mar,reim,guru,hu1,cube}, aggregation of Brownian particles \cite{cecco,pg,monoj,shm1}, noise-induced pattern formation \cite{pat0,pat,pat1}, noise induced transitions \cite{nit,nit1,nit2,denis,vitre,landa,shm2}, noise induced non equilibrium phase transitions \cite{nip0,nip,zarkin,landa1,nip1},
synchronization \cite{syn,syn1} etc. In the present paper we address an issue related to the
noise-induced transition. It has been considered in may contexts \cite{nit,nit1,nit2,denis,vitre,landa,shm2,nit3,nit4}.
A noise-induced transition occurs if
the stationary probability distribution (SPD) of the state variable is changed qualitatively as the noise intensity exceeds a critical
value. The genetic model \cite{nit1} and Hongler’s model \cite{nit2} are relevant examples demonstrating
this phenomenon. The relevant Langevin equation of motion for this kind of noise induced transition may be read as \cite{denis}

\begin{equation}
	\dot{x} =f(x)+g_1(x)\zeta_1+g_2(x)\zeta_2    \label{eq0}
\end{equation} 

\noindent
where $x$ corresponds to the coordinate of a Brownian particle which is driven by two noises,
$\zeta_1$ and $\zeta_2$, respectively. In Refs. \cite{denis,vitre}, it has been shown that for a simple case where $g_2(x)=1$ and $g_1(x)$ is nonlinear function, $g_1(x)=b\frac{x^4}{1+x^4}, b>0$ then the correlation between the two white noises may induce transition for $f(x)=-ax, a>0$ or   $f(x)=-ax^3$. The multiplicative noise induces position dependent diffusion as well as drift terms. The potential energy function due to $f(x)$ may modify at the steady state with additional fixed point(s) by virtue of the multiplicative noise. Thus the multiplicative noise may induce transition in a nonlinear system \cite{landa}. Further modification of the potential energy, as well as the transition, may happen due to the cross correlation induced drift term  \cite{denis,vitre}. 
Thus for $g_1(x)=g_2(x)=1$, noise induced transition is not possible.
It is apparent in the above examples that the multiplicative noise term (with the nonlinear function of coordinate), as well as modification of the potential energy function having new fixed point(s), is a necessary condition for the noise induced transition. But the multiplicative noise (with a linear function of coordinate) may induce transition in the nonlinear system without introducing new fixed points in the effective potential energy function at the steady state \cite{shm2}. This transition is controlled by diffusion. Thus the necessary and sufficient condition to have the noise induced transition seems to us that the total system (deterministic plus stochastic) must be a 
nonlinear one \cite{nit,nit1,nit2,denis,vitre,landa,shm2,nit3,nit4}. A similar conclusion may be drawn for another related phenomenon, noise induced non equilibrium phase transition \cite{nip0,nip,zarkin,landa1,nip1}. {\it Then one may put an open question. Is it possible to observe noise induced transition in a linear stochastic process where both deterministic and stochastic parts are linear functions of the relevant phase space variables?} To have the answer to this question we have studied the dynamics of a Brownian particle which is confined in a harmonic potential in the presence of a fluctuating magnetic field. It is an unusual type of linear stochastic process where multiplicative noises are linear functions of coordinate and velocity, respectively. This system exhibits that the noise induced transition is possible even for the case where both deterministic and stochastic parts are linear functions of the relevant phase space variables. The stationary distribution of particles in the harmonic force field is a uni-modal one at a relatively low strength of a FMF. On further increase in the strength of the field, an island may appear (having center at the origin) which is surrounded by a shallow well. It is evident from the cross section (having tri-modal) of the particle distribution in two dimension space. The cross section becomes bi-modal at a relatively large strength of the field. At this condition, the island disappears. Keeping in mind the strange behavior of the particle distribution in a linear stochastic process we have studied a stochastic nonlinear system where a Brownian particle is confined in a bi-stable potential energy field in the presence of a fluctuating magnetic field.  At low strength of the FMF the distribution of particles in space is Boltzmann type corresponding to the bi-stable potential energy field. 
On further increase in the strength of the field, an island may appear having a center at the unstable fixed point. It may be extended having $S$-shape (such that it avoids the stable fixed points) as the strength grows. At the same time, two more additional islands may appear which are located diagonally opposite to the $S$-shape island. This strange pattern may be more complex at the relatively large strength of the FMF.

Before leaving this section we would mention that the study on the dynamics of a particle in the presence of a magnetic field is always an intriguing issue in plasma physics. The field may be a stochastic one \cite{plas,plas1,eimf}. Here the motion may be non-Markovian and non-Brownian type\cite{mittal,ludger,sun}.
The fluctuating field is very relevant to give the explanation of energization of
cosmic rays and stochastic heating of plasmas\citep{mittal,shpitalnik}. 
It is to be noted here that
in recent past the magnetic field has been considered in other areas
such as barrier crossing dynamics  \cite{bag6,katsuki,pere,pereI, pereII,telang,vdo,bag3,bag4,bag5,physa}, non Markovian dynamics of a Brwonian particle in the presence of a magnetic field  \cite{bag4,bag6,physa,nmar,nmar1,nmar1I,jcp, jcpI,jcpII,he2019charge,tothova2020brownian},
stochastic thermodynamics \cite{baurapre, pal}, nonlinear dynamics \cite{nonlin} and others \cite{hsieh,hsieh1,hsieh2,indmag,indmagI,indmagII,fabio,jayn, jaynI, gelf, gelfI,ma2020nanoliquid,wang2020effect}. The present work may be relevant in the context of barrier crossing dynamics. A detailed discussion regarding this has been given in the conclusion section.

The outline of the paper is as follows: In Sec. II we have presented the model. The distribution of charged particles in space for a linear system is demonstrated in Sec.III. In the next section, we have demonstrated the distribution of charged particles in space for a nonlinear system. The paper is concluded in Sec. IV.

\section{The model}
In the present study, we have considered that a Brownian particle experiences a fluctuating magnetic field ($\textbf{B}$) along $z$-direction,
i.e., $\textbf{B}=(0,0,B(t))$. Here, $B(t)$ can be expressed as

\begin{equation}
	B(t)=B_0+B_f(t) \; \;\label{eq1}. 
\end{equation}

\noindent
Here $B_f(t)$ is the fluctuating part of the $B(t)$. It is often considered in plasma physics\cite{plas} as a signature of plasma currents where fluctuations in particle velocities and positions may occur. Magnetic
fluctuations have been measured in many tokamaks
\cite{eimf}. The fluctuating magnetic field may appear from the current fluctuations in crcuits\cite{land}. Even it may be generated by the permanent magnet \cite{ernst}. Making use of a gauge field model of the CuO planes of high-$T_c$ superconductors one may generate a FMF in a laboratory\cite{aronov}. One may generate also a fluctuating magnetic field in a laboratory by continuous optical probing
of an atomic ensemble\citep{nature}.

The Langevin equations of motion with a fluctuating magnetic field can be written as  \cite{bag6,physa,shmar}

\begin{equation}
	\dot{x} =u_x   \; \; \;, \label{eq01}
\end{equation}

\begin{equation}
	\dot{y} =u_y  \; \; \;,  \label{eq02}
\end{equation}

\begin{equation}
	m\dot{u_x} =-\frac{\partial V(x,y)}{\partial x} -\gamma_0 mu_x +(\Omega_0+\eta(t))m u_y+\frac{\dot{\eta} my}{2} + f_x(t)    \label{eq2}
\end{equation}

\noindent
and

\begin{equation}
	m\dot{u_y} = -\frac{\partial V(x,y)}{\partial y}-\gamma_0 mu_y-(\Omega_0+\eta(t)) m u_x -\frac{\dot{\eta} mx}{2} +f_y(t)    \label{eq3}
\end{equation}

\noindent
The above equations are corresponding to the motion in the $x-y$ plane. Here, $u_x$ and $u_y$ are the components of velocity of the Brownian particle with mass, $m$. $-\frac{\partial V(x,y)}{\partial x}$ and $-\frac{\partial V(x,y)}{\partial y}$ in the above equations are the components of conservative force field which is derived from the potential energy, $ V(x,y)$. $\gamma_0$ measures the strength of damping force due to thermal bath. The components of random force which is originated from the thermal bath are given by $f_x$ and $f_y$ corresponding to the motion along $x$ and $y$ directions, respectively. These are white Gaussian thermal noises, i.e.,

\begin{equation}
	\langle f_x(t)\rangle=\langle f_y(t)\rangle =0
	\label{eq4}
\end{equation}

\noindent
and
\begin{eqnarray}
	\langle f_x(t) f_x(t')\rangle=  \langle f_y(t)f_y(t')\rangle=2m \gamma_0 k_BT\delta(t-t') \; \;.
	\label{eq5}
\end{eqnarray}

\noindent
Here $k_B$ is the Boltzmann's constant and $T$ is the temperature of the thermal bath. We now consider force from the magnetic field. $\Omega_0$ and $\eta(t)$ which are associated with the magnetic force correspond to the constant and random parts of the cyclotron  frequency ($\Omega$, i.e,

\begin{equation}
	\Omega= \Omega_0+\eta  \label{eq6} \; \;,
\end{equation}

\noindent
where

\begin{equation}
	\Omega_0=\frac{qB_0}{m}  \label{eq7} 
\end{equation}

\noindent
and
\begin{equation}
	\eta(t)=\frac{qB_f(t)}{m}  \; \;\label{eq8}. 
\end{equation}

\noindent
$q$, in Eqs.(\ref{eq7}-\ref{eq8}) is the charge of the relevant Brownian particle which may experience the Lorentz force due to an applied magnetic field. Then one may identify that $\dot{\eta}$ in the force terms is due to the induced electric field and it is the time derivative of $\eta(t)$. The time evolution of the fluctuating magnetic field is assumed to be as

\begin{equation}\label{eq9}
	\dot{\eta} =  -\frac{\eta}{\tau}+\frac{\sqrt{D}}{\tau} \zeta(t) \; \;.
\end{equation}

\noindent
$D$, in the above equation, measures the strength of the fluctuating magnetic field and $\tau$ is the correlation time. $\zeta(t)$  in Eq.~(\ref{eq9}) corresponds to a white Gaussian noise having variance two. The two-time correlation function of the fluctuating field is given by

\begin{equation}
	\langle \eta(t) \eta(t') \rangle = \frac{D}{\tau} e^\frac{-|t-t'|}{\tau}  \; \;,
	\label{eq10}
\end{equation}

\noindent
Thus $\eta(t)$ is the Ornstein-Uhlenbeck  noise. It is to be noted here that in plasma physics, this kind of temporal correlation has been considered for the relevant fluctuating magnetic field\cite{ludger}. In a very recent model study, \cite{zhang} fluctuating magnetic field is considered by the Ornstein-Uhlenbeck process to 
estimate a FMF with a continuously monitored atomic ensemble. It is to be noted here that
in general the Ornstein-Uhlenbeck noise has been considered in the literature \cite{shm,majee,hu, huI,sr,bag6,shm1,shm2,syn,syn1,physa,nmar1,hang,ou, ouI, ouII, ouIII} to capture the essential feature of the non-Markovian dynamics. To avoid any confusion, we would mention here the following point.
The Ornstein-Uhlenbeck process is a Markovian one as the first order stochastic differential equation of the relevant phase space variable contains delta correlated noise. But if the 
first order differential equations of motion (which is the system's phase space description) contain
the colored noise like the Ornstein-Uhlenbeck one then the description of the Brownian motion in the system's phase space corresponds to non-Markovian dynamics. In other words, if the phase space is extended including the first order stochastic differential equation (with white noise) for the time evolution of the 
colored noise then the stochastic process in extended phase space is a Markovian one. Thus in the extended phase space, all the Brownian motions are Markovian dynamics. Then the classification of Markovian and non Markovian dynamics is based on the description of the motion in terms of the system's phase space variables. 
Thus the present system corresponds to the non Markovian dynamics as the equations of motion in terms of system's phase space variable contain the colored noise. Therefore in the present manuscript, one may call the dynamics of the system the non Markovian one.

To describe the diffusive behavior in plasma there are different kinds of Langevin equations of motion. One may consider the equations of motion like Eqs. (\ref{eq2}-\ref{eq3}) where the deterministic force corresponds to the average force field (which may be a function of both coordinate and velocity) and the random force is due to the collisions\cite{lewcm}. In some cases, the Langevin equations of motion are assumed to be with magnetic force only\cite{ludger,sun}  or magnetic force plus collisional random force\cite{spineanu}. The relevant Langevin equation of motion may be such that it corresponds to the Brownian motion of a free particle in the presence of a fluctuating magnetic field\cite{plas1}.
To explain the energization of cosmic rays one may assume the equations of motion having forces from the fluctuating magnetic field and the associated induced electric field\cite{mittal}. 
Finally, to describe the motion of an electron 
in a semiconductor or ion in electrolytes Eqs.~(\ref{eq2}-\ref{eq3}) may be useful. Here the relevant Brownian particle is supposed to experience a conservative force from the potential energy field. One
may assume that the energy field is periodic in space with a finite energy barrier between two consecutive wells.
At the bottom of the well, the potential energy field seems to be a harmonic one. Then we often say that
the study on the harmonic oscillator is not a mere example. However,
in the next section, we will explore how the distribution of particles (in a harmonic potential) depends on the fluctuating magnetic field. 
Then to identify the signature of a nonlinear potential energy field in this context, one may consider the energy function with double well. It may correspond to two consecutive wells of a periodic potential energy field. At the same time, one may get the advantage to avoid the problem of normalization which requires confining the particle in finite space. In Sec. IV we will explore the distribution of particles in a double well potential energy field in the presence of a FMF. 
Thus Eqs.(\ref{eq2}-\ref{eq3}) may be useful at different contexts. 
It is to be noted here that in the recent
past, harmonic or anharmonic oscillator in the presence of a magnetic field has
been considered in different contexts 
\cite{bag6,katsuki,pere,pereI, pereII,telang,vdo,bag3,bag4,bag5,physa,bag4,physa,nmar,nmar1,nmar1I,jcp, jcpI,jcpII,he2019charge,tothova2020brownian,baurapre, pal,nonlin,hsieh,hsieh1,hsieh2,indmag,indmagI,indmagII,fabio,jayn, jaynI, gelf, gelfI,ma2020nanoliquid,wang2020effect}.

Before leaving this equation we would mention the following pertinent point. Eqs. (\ref{eq2}-\ref{eq3}) seems to be an incomplete description of the motion since these do not include the effect from the induced magnetic field due to the time dependent electric field as suggested by the Maxwell's equation. One may easily check here that the induced electric field which appears in the equation of the motion is related to the applied magnetic field by

\begin{equation}
	\nabla \times \textbf{E}_{ind}=-\frac{\partial }{\partial t}\textbf{B} =-\dot{\eta}(t) \; \; \;, \label{indel1}
\end{equation}

\noindent
where $\textbf{E}_{ind}(=\left(\frac{\dot{\eta}(t) y}{2},-\frac{\dot{\eta}(t) x}{2},0\right))$ is the induced electric field. The Maxwell's equation with this electric field for the present system can be read as

\begin{equation}
	\nabla \times \textbf{B}_{ind}=\mu\epsilon\frac{\partial }{\partial t}\textbf{E}_{ind}  \; \; \;. \label{indmf}
\end{equation}

\noindent
Here $\textbf{B}_{ind}$ is the induced magnetic field due to the time dependent induced electric field.
The applied magnetic field does not appear in the above equation since its curl is zero. $\epsilon$ and $\mu$ correspond to the permittivity and
the permeability of the electrolytic medium. Taking curl in both sides of the above equation and then using Eq.(\ref{indel1}) and  
$\nabla \bf{\cdot} \textbf{B}_{ind}=0$ into this we have

\begin{equation}
	\frac{\partial^2 }{\partial^2 z}B_{indz}=\mu\epsilon\ddot{\eta}(t)  \; \; \;, \label{indmf1}
\end{equation} 

\noindent
where $B_{indz}$ is the $z$-component of the induced magnetic field. The solution of the above equation can be read as

\begin{equation}
	B_{indz}=\mu\epsilon\ddot{\eta}(t)z^2 \; \; \;. \label{indmf2}
\end{equation} 

\noindent
Here we have used $\frac{\partial }{\partial z}B_{indz}=0$ at $t=0$ and $z=0$. It is to be noted here that 
the magnitude of the product of $\mu$ and $\epsilon$ may be of the order of $10^{-17} (m/sec)^{-2}$. Then above equation certainly implies that the induced magnetic field is negligible compared to the applied magnetic field. This might be the reason to exclude the effect from the induced magnetic field in the equation of motion in the earlier studies\cite{bag5,bag6,physa,ludger,mittal,saha}. 

\begin{figure}[t!]
	\centering	
	\includegraphics[width=1.0\columnwidth,angle=0,clip]{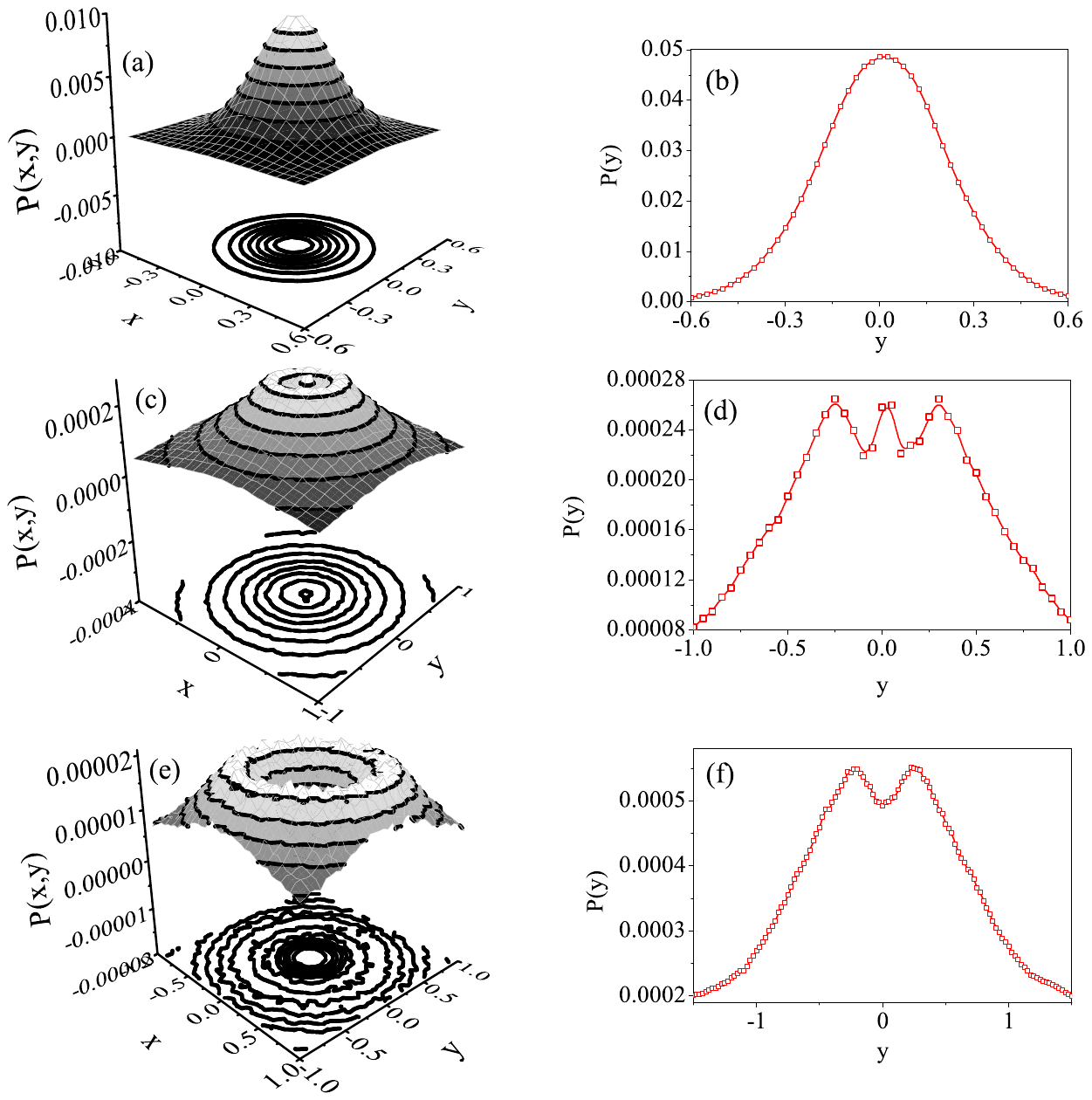}
	\caption{Demonstration of distribution of charged particles in space for a linear system for the parameter set, $\omega^2=2, \Omega_0=7.0$, $\tau=0.2, \gamma=0.1$ and $k_BT=0.025$. (a) Plot of reduce distribution function ($P(x,y)$ {\it vs} coordinate for $D=0.04$. (b) Its cross section along $x=0$. (c)  Plot of reduce distribution function ($P(x,y)$ {\it vs} coordinate for $D=0.057$(d) Its cross section along $x=0$. (e)  Plot of reduce distribution function ($P(x,y)$ {\it vs} coordinate $D=0.0575$ (f) Its cross section along $x=0$.  (Units are arbitrary)}
	\label{fig1}
\end{figure}

 \section{Noise induced transition in a fluctuating magnetic field driven harmonic oscillator: a special signature of the field with a new mechanism}
 
 In this section, we consider that the Brownian motion is confined in the two dimensional harmonic potentials with the angular frequency $\omega$ and it can be read as
 
 \begin{equation}\label{eq11}
 	V(x,y)=\omega^2m(x^2+y^2)/2 \; \; \;.
 \end{equation}
 
 \noindent
 Using Eq. (\ref{eq11}) into the Eqs. (\ref{eq2}-\ref{eq3}) we have
 
 \begin{equation}\label{eq12}
 	\dot{u_x} =-\omega^2 x -\gamma_0 u_x +(\Omega_0+\eta(t)) u_y+\frac{\dot{\eta} y}{2} + f_x(t)  
 \end{equation}
 
 \noindent
 and
 
 \begin{equation}\label{eq13}
 	\dot{u_y} = -\omega^2 y-\gamma_0 u_y-(\Omega_0+\eta(t))  u_x-\frac{\dot{\eta} x}{2}+f_y(t) \; \; \;.   
 \end{equation}

 \begin{figure}[t!]
 	\centering
 	\includegraphics[width=1.0\columnwidth,angle=0,clip]{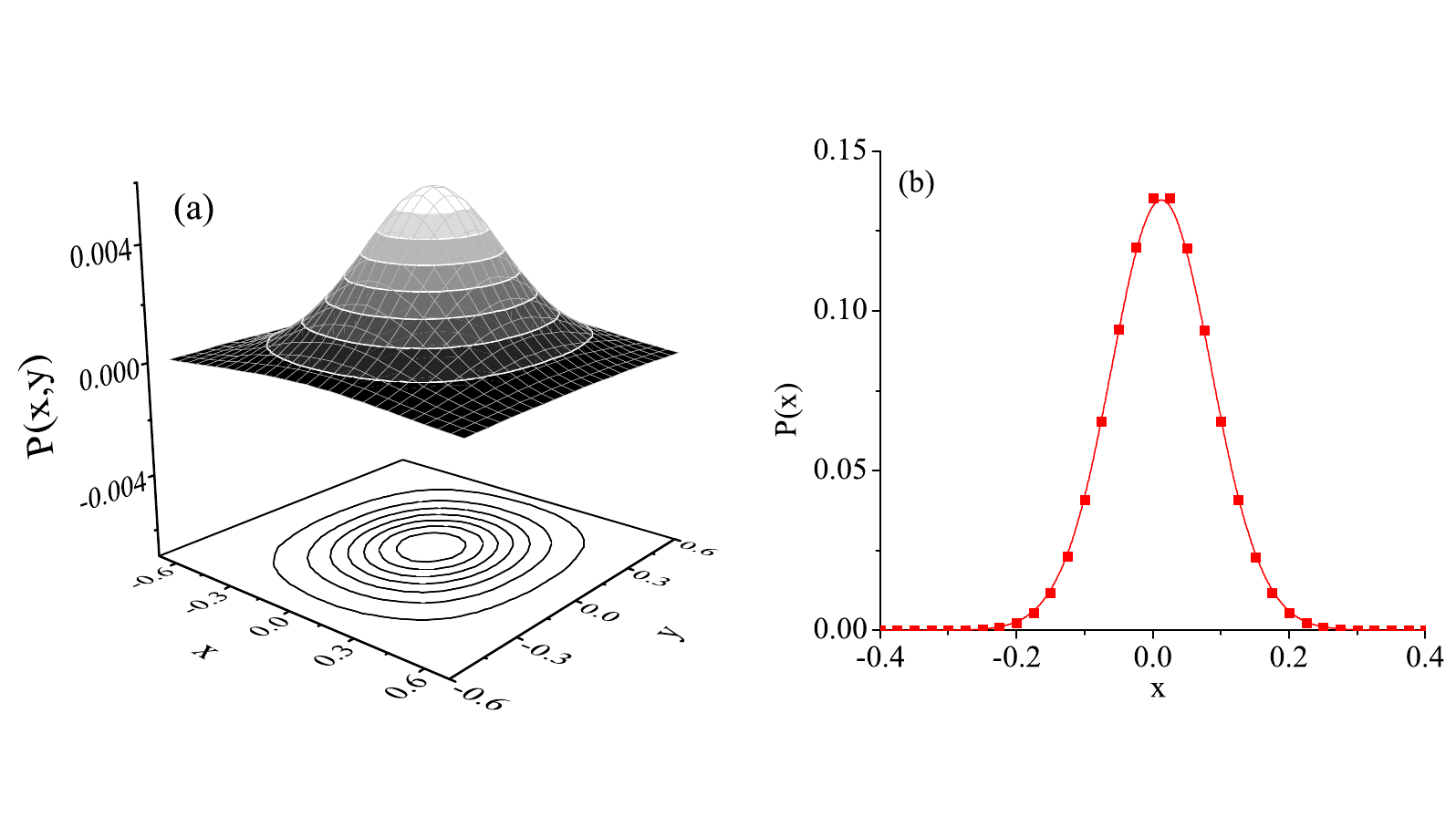}
 	\caption{(a) Demonstration of distribution of particles for the linear system with equations of motion (\ref{ad1}-\ref{ad2}) in space for the parameter set, $\omega^2=2.0$, $\tau=0.2, \gamma=0.1$, $D=0.1$ and $k_B T=0.1$. (b) Demonstration of distribution of particles for the linear system with equation of motion (\ref{ad3}) in space for the same parameter set. 
 		(Units are arbitrary)}
 	\label{fig2}
 \end{figure}
 
 \noindent
 In the Eqs.(\ref{eq12}-\ref{eq13}) we have used $m=1$. We will follow this in the rest of the part also. 
 However, the fluctuating magnetic field induces an unusual type of multiplicative noise. It is multiplied by velocity and coordinate, respectively. 
 In the usual multiplicative noise driven process \cite{denis,vitre,landa,shm2}, the noise is multiplied with the coordinate along which the Brownian particle experiences a random force. However, the above equations can be decoupled as
 \begin{equation}
 \begin{array}{rcl}
 \dot{u}_{x} & = & -\left[\omega^{2}+\left(\Omega_{0}+\eta(t)\right)\dfrac{d}{dt}\left(L^{-1}\dfrac{\dot{\eta}}{2}\right)+\dfrac{\dot{\eta}}{2}L^{-1}\left(\dfrac{\dot{\eta}}{2}\right)\right]x\\
 &  & -\left[\gamma_{0}+\left(\Omega_{0}+\eta(t)\right)\dfrac{d}{dt}\left(L^{-1}\left\{ \Omega_{0}+\eta(t)\right\} \right)+\dfrac{\dot{\eta}}{2}L^{-1}\left\{ \Omega_{0}+\eta(t)\right\} \right]u_{x}\\
 &  & +\left(\Omega_{0}+\eta(t)\right)\left[\dfrac{d}{dt}\left(e^{-Kt}\alpha_{1}(t)\right)+\dfrac{d}{dt}\left(L^{-1}\left\{ f_{y}\left(t\right)\right\} \right)\right]+\dfrac{\dot{\eta}}{2}\left[e^{-Kt    }\alpha_{1}(t)+L^{-1}\left\{ f_{y}\left(t\right)\right\} \right]+f_{x}(t)
 \end{array}\label{bn1}
 \end{equation}
 
 \noindent
 and
 
 \begin{equation}
 \begin{array}{rcl}
 \dot{u}_{y} & = & -\left[\omega^{2}+\left(\Omega_{0}+\eta(t)\right)\dfrac{d}{dt}\left(L^{-1}\dfrac{\dot{\eta}}{2}\right)+\dfrac{\dot{\eta}}{2}L^{-1}\left(\dfrac{\dot{\eta}}{2}\right)\right]y\\
 &  & -\left[\gamma_{0}+\left(\Omega_{0}+\eta(t)\right)\dfrac{d}{dt}\left(L^{-1}\left\{ \Omega_{0}+\eta(t)\right\} \right)+\dfrac{\dot{\eta}}{2}L^{-1}\left\{ \Omega_{0}+\eta(t)\right\} \right]u_{y}\\
 &  & -\left(\Omega_{0}+\eta(t)\right)\left[\dfrac{d}{dt}\left(e^{-Kt}\alpha_{1}(t)\right)+\dfrac{d}{dt}\left(L^{-1}\left\{ f_{x}(t)\right\} \right)\right]-\dfrac{\dot{\eta}}{2}\left[e^{-Kt}\alpha_{1}    (t)+L^{-1}\left\{ f_{x}\left(t\right)\right\} \right]+f_{y}(t) \; \; \;.
 \end{array}\label{bn2}
 \end{equation}

 \noindent
 Derivation of these equations has been given in Appendix. Then it is to be noted here that Eq.(29) in Ref.\cite{bag6} is not an exact one.
 But the results in this reference are not affected as the equation was not used in the calculation.
 
 It is difficult to solve both the coupled and the decoupled equations of motion even for the linear system. It is also difficult to write the Fokker-Planck equation in the phase space. Therefore we are restricted to studying the present problem numerically. We have solved the Eqs.(\ref{eq12}-\ref{eq13}) using the Heun's
 method \cite{rt}. It is a stochastic version of the Euler method which reduces to
 the second order Runge-Kutta method in the absence of noise. Details as well reliability of the method are described in Refs. \cite{physa}, and \cite{bag6}, respectively.
 Based on this method we have calculated stationary reduced distribution function, $P(x,y)$ for different cases and demonstrated in Fig.~1. It shows that with an increase in strength of the fluctuating magnetic field the probability distribution function may be folded in space even (as shown in panel (c)) for the linear system. 
 The cross section at $x=0$ of the distribution of particles in space is plotted in panel (d) which explicitly demonstrates the non monotonic change in the distribution of particles in space. The panel (e) is also an interesting one where the distribution function is exhibited at a relatively large strength of the fluctuating magnetic field. Here a shallow basin with a minimum at the origin appears which is surrounded by an island. This is explicit in panel (f) which is the cross section of the panel (e) at $x=0$. Of course one may expect a transition from panel (c) to (e) on further increase in strength of the fluctuating magnetic field. If the noise strength is appreciably large then the distribution function seems to be not a smooth and continuous one which may correspond to a composed of bound and unbound motions. However, to check whether the panels like (c) and (e) are the special signatures of the fluctuating field or not we consider dynamics with the following equations of motion,
 
 \begin{equation}\label{ad1}
 	\dot{u_x} =-\omega^2 x -\gamma_0 u_x +\eta y + f_x(t)  
 \end{equation}
 
 \noindent
 and
 
 \begin{equation}\label{ad2}
 	\dot{u_y} = -\omega^2 y-\gamma_0 u_y +\eta x+f_y(t) \; \; \;.   
 \end{equation}

 \noindent
 For this dynamics, the distribution of particles in space has been shown in panel (a) of Fig.~2. It shows that the probability of finding the Brownian particle decreases monotonically with an increase in distance from the origin. We have checked that this pattern does not change even at a relatively high strength of the noise. In other words, for this system noise induced transition is not possible. We now consider another case with the following equation of motion

\begin{equation}\label{ad3}
 	\dot{u_x} =-\omega^2 x -\gamma_0 u_x +\eta x + f_x(t)  
 \end{equation}
 
 \noindent
 The probability of particle for this case also decreases monotonically with an increase in distance from the origin as shown in panel (b) of Fig.~2. We have checked that this pattern does not change even at a relatively high strength of the noise. In other words, for this system noise induced transition is not possible. Thus Fig.~1 and Fig.~2 clearly suggest that the non monotonic change of distribution of charged particles in space for the linear stochastic system (where both the deterministic and the stochastic parts are linear functions of the system's phase space variables) may be a special signature of the fluctuating magnetic field. 
 
 \begin{figure}[t!]
 	\centering
 	\includegraphics[width=1.0\columnwidth,angle=0,clip]{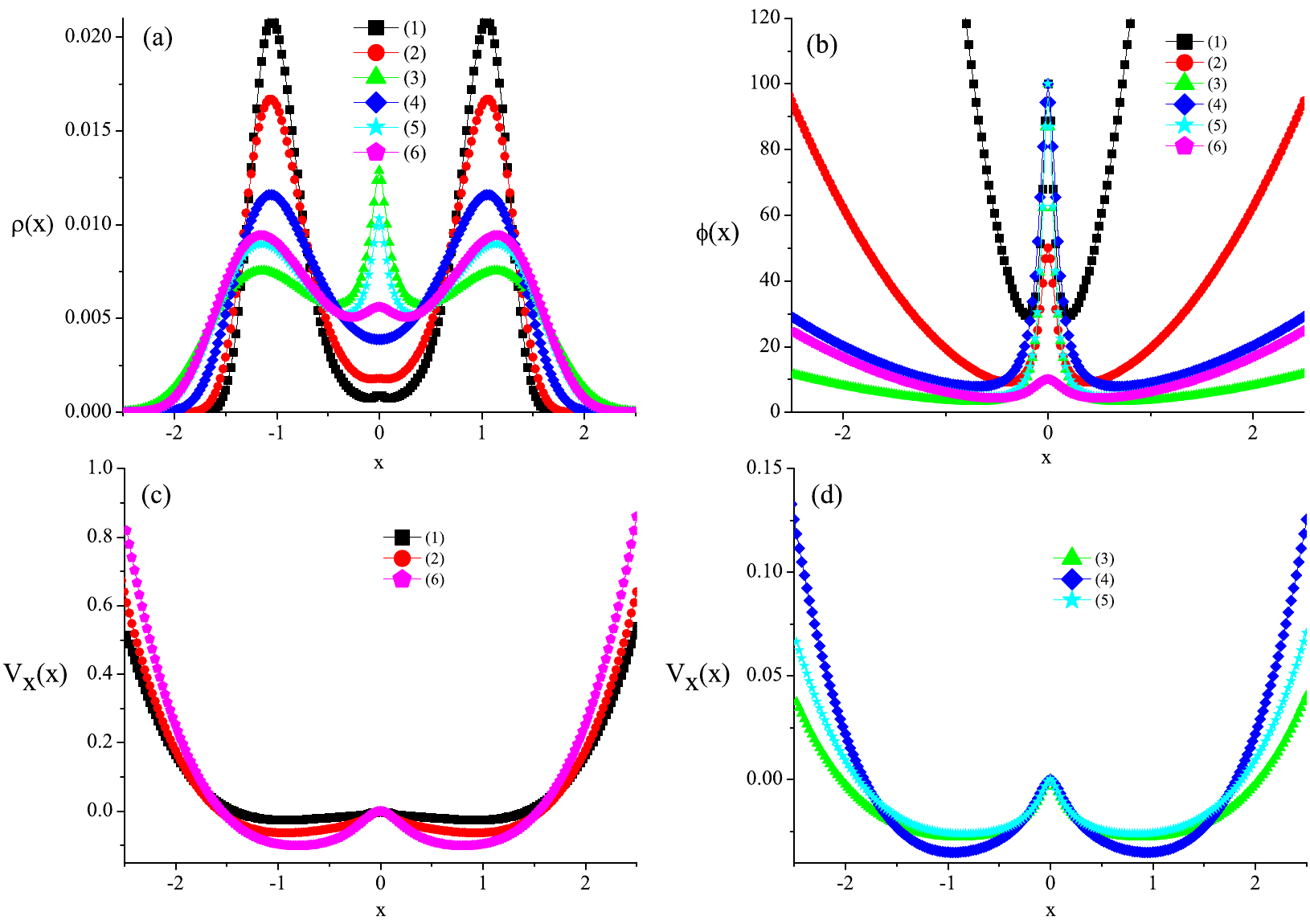}
 	\caption{ (a) Plot of the reduced distribution function ($\rho(x)$) vs coordinate for the parameter set  $a = 0.25$; $b = 0.5$; and (1) $ k_BT=0.01, \gamma= 1.0, D=2.5, \tau=10.0$ (2) $ k_BT=0.01, \gamma= 0.5, D=2.5, \tau=1.5$. (3) $ k_BT=0.01, \gamma= 1.0, D=2.5, \tau=1.0$ (4) $ k_BT=0.01, \gamma= 1.0, D=1.0, \tau=1.0$ (5) $ k_BT=0.01, \gamma= 1.0, D=2.5, \tau=1.5$ (6) $ k_BT=0.1, \gamma= 1.0, D=2.5, \tau=1.5$. (b) Plot of $\rho$ vs coordinate corresponding to panel (a). (c,d) Plot of the renormalized potential energy function $(V_x(x))$ vs coordinate corresponding to panel (a). (Units are arbitrary)}
 	\label{fig3}
 \end{figure} 
 
 \subsection{Understanding about the non monotonic change of distribution of particles in a harmonic potential well}
 
 Panels (c) and (e) in Fig.~1 clearly imply that the distribution of particles at the steady state in the presence of a fluctuating magnetic field may be drastically different from the Boltzmann type. Then
 one may think that the FMF induced transition is due to a change of potential energy field with additional fixed point(s) as happens in the case of noise induced transition. But it is very difficult to determine the effective potential energy field at the stationary state for the present system as mentioned above. Then
 one may invoke a recent Ref.~\cite{shm2} to have an idea of whether the modification of the potential energy function with new fixed point(s) is possible or not in the presence of the fluctuating magnetic field.  The relevant equation of motion in Ref.~\cite{shm2} 
 is
 
 \begin{equation}
 	\dot{x} =f(x)+ x \eta_1(t)/\gamma +x\eta_2(t)/\gamma+ f_x(t)/\gamma   \; \; \;\label{eq18a}
 \end{equation}
 
 \noindent
 with
 \begin{equation}
 	f(x)=-\frac{1}{\gamma}\frac{dV(x)}{dx} \; \;. \label{eq18fa}
 \end{equation}
 \noindent
 It is a description of over damped Brownian motion in the presence of external two multiplicative noises
 , $\eta_1(t)$ and $\eta_2(t)$, respectively. Their statistical properties are defined as
 
 \begin{equation}
 	\langle \eta_1(t) \rangle  = \langle \eta_1(t) \rangle=0  \; \;,
 	\label{eqcf}
 \end{equation}
 
 \begin{equation}
 	\langle \eta_1(t) \eta_1(t') \rangle = \frac{D_1}{\tau_1} e^\frac{-|t-t'|}{\tau_1}  \; \;,
 	\label{eqcf1}
 \end{equation}

 \begin{equation}
 	\langle \eta_2(t) \eta_2(t') \rangle = \frac{D_2}{\tau_2} e^\frac{-|t-t'|}{\tau_2}  \; \;,
 	\label{eqcf2}
 \end{equation}
 
 \noindent
 and
 \begin{equation}
 	\langle \eta_1(t) \eta_2(t') \rangle = \langle \eta_2(t) \eta_1(t')\rangle=\frac{\lambda\sqrt{D_1D_2}}{\tau_c} e^\frac{-|t-t'|}{\tau_c}  \; \;.
 	\label{eqcf3}
 \end{equation}
 \noindent
 Here $\lambda$ measures the strength of cross correlation between the noises , $\eta_1(t)$ and $\eta_2(t)$ with the correlation time $\tau_c$. Then the Langevin Eq.(\ref{eq18a}) can be  expressed effectively as
 \begin{equation}
 	\dot{x} =f(x)+x\eta_e(t) + f_x(t)/\gamma   \; \; \;. \label{eq18b}
 \end{equation}
 with 
 \begin{equation}
 	\eta_e(t)=(\eta_1+\eta_2)/\gamma \; \; \;, \label{eqcf4}
 \end{equation}
 and
 \begin{equation}
 	\langle \eta_e(t) \eta_e(t') \rangle = \frac{D_e}{\tau_e} e^\frac{-|t-t'|}{\tau_e}  \; \;.
 	\label{eqcf5}
 \end{equation}
 \noindent
 Here $D_e$ and $\tau_e$ are defined as

 \begin{equation}
 	D_e=\int_0^\infty\langle \eta_e(t) \eta_e(0) \rangle dt  
 	\label{eqcf6}
 \end{equation}
 
 \noindent
 and
 
 \begin{equation}
 	\tau_e=\int_0^\infty t \langle \eta_e(t) \eta_e(0) \rangle dt
 	\label{eqcf7}
 \end{equation}
 
 \noindent
 Then using Eq.(\ref{eqcf4}) into Eqs. (\ref{eqcf6}-\ref{eqcf7}) one can write that
 
 \begin{equation}
 	D_e=\frac{D_1+2\lambda\sqrt{D_1D_2}+D_2}{\gamma^2}
 	\label{eqcf8}
 \end{equation}
 
 \noindent
 and
 
 \begin{equation}
 	\tau_e=\frac{D_1\tau_1+2\lambda\sqrt{D_1D_2}\tau_c+D_2\tau_2}{D_e\gamma^2}
 	\label{eqcf9}
 \end{equation}
 
 \noindent
 Thus $\eta_e(t)$ in Eq.(\ref{eq18b}) is an Ornstein-Uhlenbeck  noise like $\eta(t)$. 
 
 The stationary distribution function for the equation (\ref{eq18b}) of motion can be read as \cite{shm2}
 
 \begin{equation}
 	\rho (x)= N\phi(x)\exp[-\frac{V_x(x)}{D_T}] \; \; \;,
 	\label{eq18c}
 \end{equation}
 
 \noindent
 where
 
 \begin{equation}
 	\phi(x) = \frac{A(x)^2}{(D_ex^2+D_T)}\; \; \;,
 	\label{eq18d}
 \end{equation}
 
 \noindent
 and
 
 \begin{equation}
 	V_x(x)=-\int_0^x\frac{(f(x')A(x')+D_e x')A(x')-A'(x')(D_eq'^2+D_T)}{A(x')(\frac{D_ex'^2}{D_T}+1)}dx' \; \; \;.
 	\label{eq18e}
 \end{equation}

 \noindent
 with 
 \begin{equation}
 	A(x)=1-\tau[f'(x)-f(x)/x] \; \; \label{eq18f}
 \end{equation}
 
 \noindent
 and
 
 \begin{equation}
 	D_T=k_B T/\gamma \; \;. \label{eq18g}
 \end{equation}
 
 \noindent
 $\phi(x)$ and $V_x(x)$ are interpreted in Ref.~\cite{shm2} as the inverse of position dependent diffusion coefficient and the re-normalized potential energy function, respectively. Making use of the equation, $\frac{d\rho(x)}{dx}=0$ which defines the location of the extrema of $\rho(x)$, in the form
 
 \begin{equation}
 	A^\prime(x)[D_ex^2+D_T]-xD_eA(x)+f(x)A(x)^2=0 \; \;. \label{eq18h}
 \end{equation} 
 
 \noindent
 where $A^\prime(x)=\frac{dA(x)}{dx}$. Similarly one may define the location of the extrema of the re-normalized potential energy field, $V_x(x)$, in the form

 \begin{equation}
 	A(x)[D_ex+f(x)A(x)]-A^\prime(x)[D_ex^2+D_T]=0 \; \;. \label{eq18hi}
 \end{equation}
 
 \noindent
 Now, one may check easily that for a harmonic potential energy field, $V(x)=\omega^2 x^2/2$, 
 $\rho(x)$ is a uni-modal one with a maximum at $x=0$ and the re-normalized potential energy field (RNPEF) mimics $V(x)$. Thus in the presence of a harmonic force field, the multiplicative noise which is a linear function of coordinate can not introduce an additional fixed point into the RNPEF. Since the noise induced drift term is a linear one, it may modulate the harmonic frequency of the renormalized potential energy field but can not create a new fixed point. Thus the noise induced transition is not possible for the linear systems (\ref{ad1}-\ref{ad2}) and (\ref{ad3}), respectively. Similarly this phenomenon may not appear for the equation (\ref{eq0}) of motion if $f(x)$, $g_1(x)$ and $g_2(x)$ are linear functions of $x$.
 
 For further check one may consider a bistable potential energy field, $V(x)=a x^4 -b x^2$. $a, b$  are constants that determine the location of the extrema of the field. Then from Eqs.(\ref{eq18h}-\ref{eq18hi}) we
 have

 \begin{equation}
 	128\frac{a^3\tau^2}{\gamma^2}x^7+32\frac{a\tau}{\gamma}(a-2\frac{ab\tau}{\gamma})x^5+
 	+2a(a-8\frac{b\tau}{\gamma}+6\tau D_e)x^3+(8a\tau D_T-b+D_e)x=0 \; \;. \label{eq18hj}
 \end{equation}   
 
 \noindent
 and
 
 \begin{equation}
 	128\frac{a^3\tau^2}{\gamma^2}x^7+32\frac{a\tau}{\gamma}(a-2\frac{ab\tau}{\gamma})x^5+
 	+2a(a-8\frac{b\tau}{\gamma}+4\tau D_e)x^3+(8a\tau D_T-b)x=0 \; \;. \label{eq18hk}
 \end{equation}  
 
 \begin{figure}[t!]
 	\centering
 	\includegraphics[width=1.0\columnwidth,angle=0,clip]{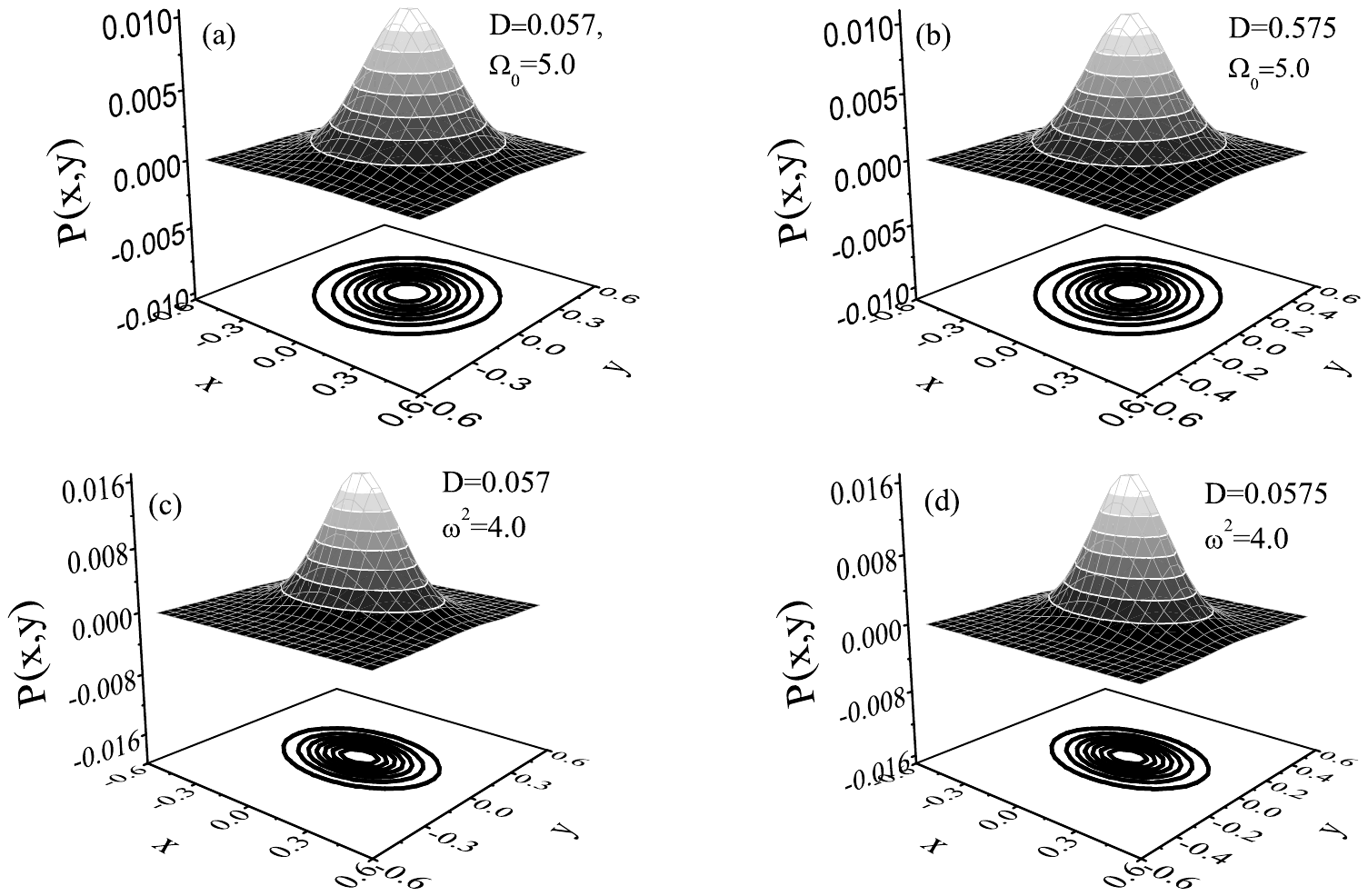}
 	\vspace{-0.1in}
 	\caption{Demonstration of distribution of charged particles in space for a linear system for the parameter set,  $\tau=0.2, \gamma=0.1$ and $k_BT=0.025$. (a, b) Plot of reduce distribution function ($P(x,y)$ {\it vs} coordinate for $\omega^2=2$. (c,d)  Plot of reduce distribution function ($P(x,y)$ {\it vs} coordinate for $\Omega_0=7.0$. (Units are arbitrary)}
 	\label{fig4}
 	\vspace{-0.1in}
 \end{figure}

\noindent 
$x=0$ is a common root for both cases. It is difficult to determine other roots of the above equations. Then to make the presentation self-sufficient we demonstrate the distribution function and the related quantities in Fig.~3. Thus five real roots may be possible corresponding to locations of extrema of the distribution function for a given parameter set. But for the same parameter set three extrema to appear for the re-normalized potential energy field as like as $V(x)$. Thus again it is proved that the multiplicative noise which is a linear function of coordinate can not introduce additional fixed points into the RNPEF. It is to be noted here that the locations of extrema which are nearby to the origin for the distribution function (with the potential energy, $V(x)=a x^4 -b x^2$) have no correspondence with the renormalized potential energy field. In other words, the locations of extrema that are far away from the origin of the distribution function have correspondence with the RNPEF. These can be read
approximately from Eqs.~(\ref{eq18hj}-\ref{eq18hk}) as 

\begin{equation}
	x_{\pm}\simeq\pm\sqrt{\frac{b}{2a}-\frac{\gamma}{4a\tau}} \; \;. \label{eq18hl}
\end{equation}

Thus the noise induced transition is controlled by the position dependent diffusion. The unexpected maximum at the origin for the distribution function may be due to the very weak diffusion at this region. The apparent unstable motion in between stable and unstable fixed points of the potential energy field is a balance like situation where the force from the potential energy favors to find the particle but the position dependent diffusion opposes it.  

Then from the above discussion, it is apparent that the fluctuating magnetic field induced transition in a linear stochastic system seems to be due to implicit dynamics. Since the terms in equations of motion with multiplicative noises are a linear function of phase space variables then the drift terms which are introduced by these can not create new fixed points in the re-normalized potential energy field. It may be irrespective of the nature of the multiplicative noises. For further details, one may go through Ref.~\cite{shm2}.    

We are now in a position to explore the relevant implicit dynamics (if any). 
Fig.~2 implies that the time independent part of the applied magnetic field has an important role in the manifestation of the unique signature. Then we demonstrate the fate of panels (c) and (e) in Fig.~1 at a relatively low value of $\Omega_0$ in Fig.~4. On the other hand, If the strength of the time independent magnetic field is appreciably large then the distribution function seems to be not a smooth and continuous one as happens in the case of fluctuating magnetic field. However, to know the probable reason which is related to the present context we consider the following equations of motion   

\begin{equation}
	\dot{u}_x = -\omega^2 x  + \Omega_0 u_y \label{eq33d}
\end{equation}

\noindent
and

\begin{equation}
	\dot{u}_y = -\omega^2 y-\Omega_0 u_x   \; \; \;. \label{eq34d}
\end{equation}

\noindent
The above coupled equations of motion can be solved using the transformation, $\xi=x+iy$\cite{aquino1,landau}. Then we have

\begin{equation}
	\ddot{\xi} = -\omega^2 \xi -2\beta\dot{\xi} \label{eqzeta} \; \; \;,
\end{equation}

\noindent
where $\beta=i\Omega_0/2$. This leads to having the solution of the above equation as

\begin{equation}
	\xi(t) = \xi(0)e^{-\beta t} cos(\sqrt{\omega^2-\beta^2}t) \label{eqzeta1} \; \; \;.
\end{equation}

\noindent
Decomposing it time dependence of $x(t)$ and $y(t)$ can be read as

\begin{eqnarray}
	x(t) & = & \frac{x(0)}{2}[cos(\sqrt{\omega^2+\Omega_0^2/4}+\Omega_0/2)t+cos(\sqrt{\omega^2+\Omega_0^2/4}-\Omega_0/2)t]\nonumber\\
	&+&\frac{y(0)}{2}[sin(\sqrt{\omega^2+\Omega_0^2/4}+\Omega_0/2)t-sin(\sqrt{\omega^2+\Omega_0^2/4}-\Omega_0/2)t]
	\label{eqzeta4} \; \; \;
\end{eqnarray}

\noindent
and

\begin{eqnarray}
	y(t) & = & \frac{y(0)}{2}[cos(\sqrt{\omega^2+\Omega_0^2/4}+\Omega_0/2)t+cos(\sqrt{\omega^2+\Omega_0^2/4}-\Omega_0/2)t]\nonumber\\
	&-&\frac{x(0)}{2}[sin(\sqrt{\omega^2+\Omega_0^2/4}+\Omega_0/2)t-sin(\sqrt{\omega^2+\Omega_0^2/4}-\Omega_0)/2)t]
	\label{eqzeta5} \; \; \;.
\end{eqnarray}

\begin{figure}[t!]
	\centering
	\includegraphics[width=1.0\columnwidth,angle=0,clip]{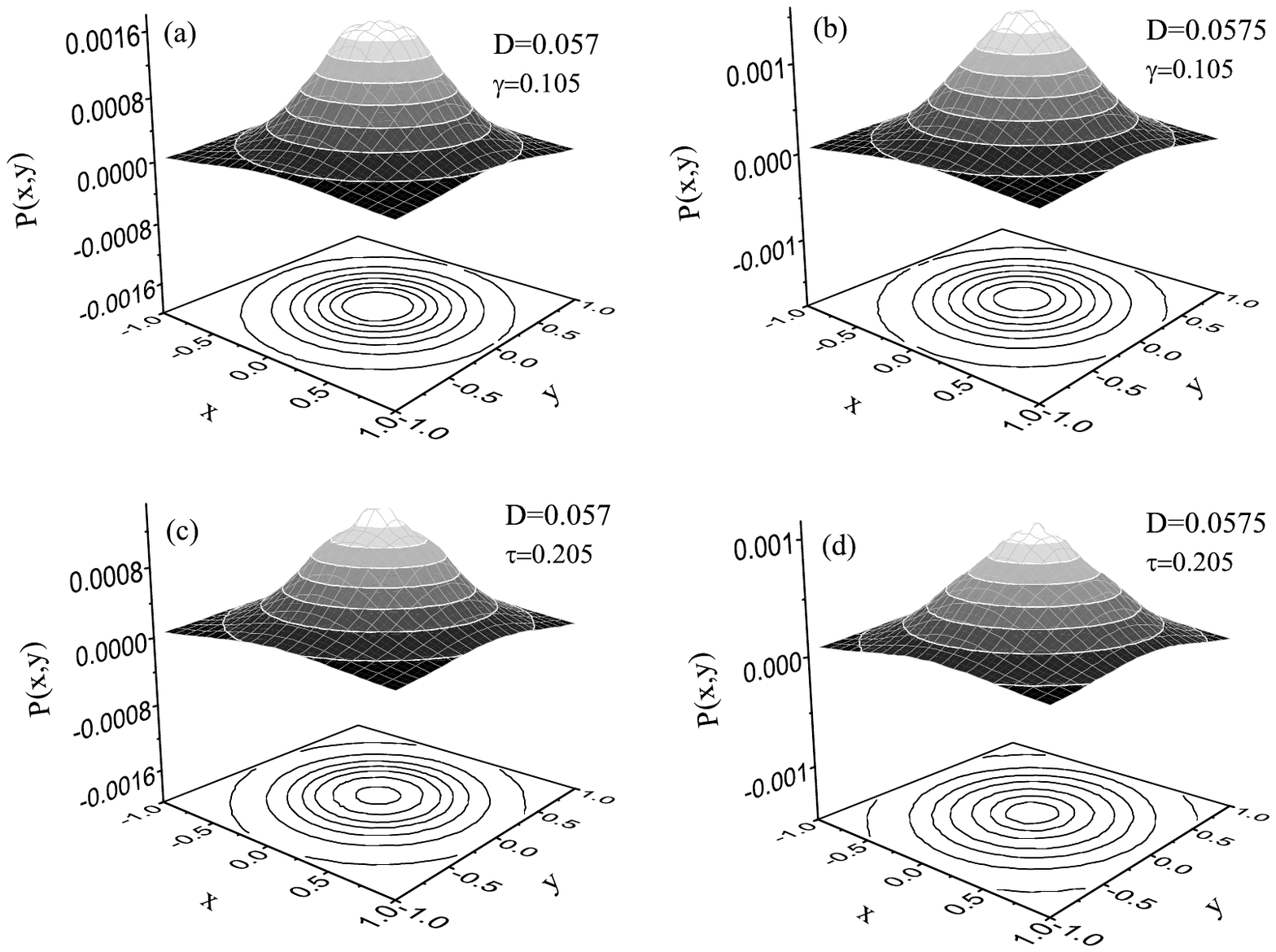}
	\caption{Demonstration of distribution of charged particles in space for a linear system for the parameter set, $\omega^2=2,\Omega_0=7.0$ and $k_BT=0.025$. (a,b) Plot of reduce distribution function ($P(x,y)$ {\it vs} coordinate for $\tau=0.2$. (c,d)  Plot of reduce distribution function ($P(x,y)$ {\it vs} coordinate for $\gamma=0.1$. (Units are arbitrary)}
	\label{fig5}
\end{figure}

\noindent
Now one can check easily that the above relations reduce to the expected results i.e., the simple harmonic motion at the limit $\Omega_0=0.0$. But in the presence of a magnetic field 
$x(t)$ as well as $y(t)$ are composed of two vibrational modes with angular frequencies, $\omega_1=\sqrt{\omega^2+\Omega_0^2/4}+\Omega_0/2$ and $\omega_2=\sqrt{\omega^2+\Omega_0^2/4}-\Omega_0/2$ respectively. Thus if the field strength is sufficiently large such that the harmonic force seems to be very weak then the high frequency mode approaches a cyclotron motion and the other one may behave as like a free particle one. Both the modes may be important to appear the noise induced transition in a linear stochastic system.  The low frequency mode makes the induced electric field  significant at a relatively large strength of the fluctuating magnetic field. Then the motion at a steady state seems to be like a cyclotron one as implied by the high frequency mode. It might be apparent in panel (e) of Fig.~1.
In other words, if the fluctuating magnetic, as well as induced electric fields, are weak then the stationary distribution of particles may be close to the equilibrium one with a maximum at the origin (as shown in panel (a) of Fig.~1) corresponding to a stable fixed point. An interim situation may be like panel (c) in the same figure. This explanation implies that at relatively low strength of the time 
independent magnetic field or high frequency of the harmonic oscillator, the distribution of particles may be like the Boltzmann type. 
Thus Fig.~4 might be the justification of the explanation. At the same time, one may not expect the noise induced transition for dynamics which are described by Eqs. (\ref{ad1}-\ref{ad2}) and (\ref{ad3}), respectively. Thus the noise induced transition in a linear stochastic system (where both the deterministic and the stochastic parts are linear functions of system's
phase space variables) may be a unique signature of the magnetic field with a new mechanism.

\begin{figure}[t!]
	\centering
	\includegraphics[width=1.0\columnwidth,angle=0,clip]{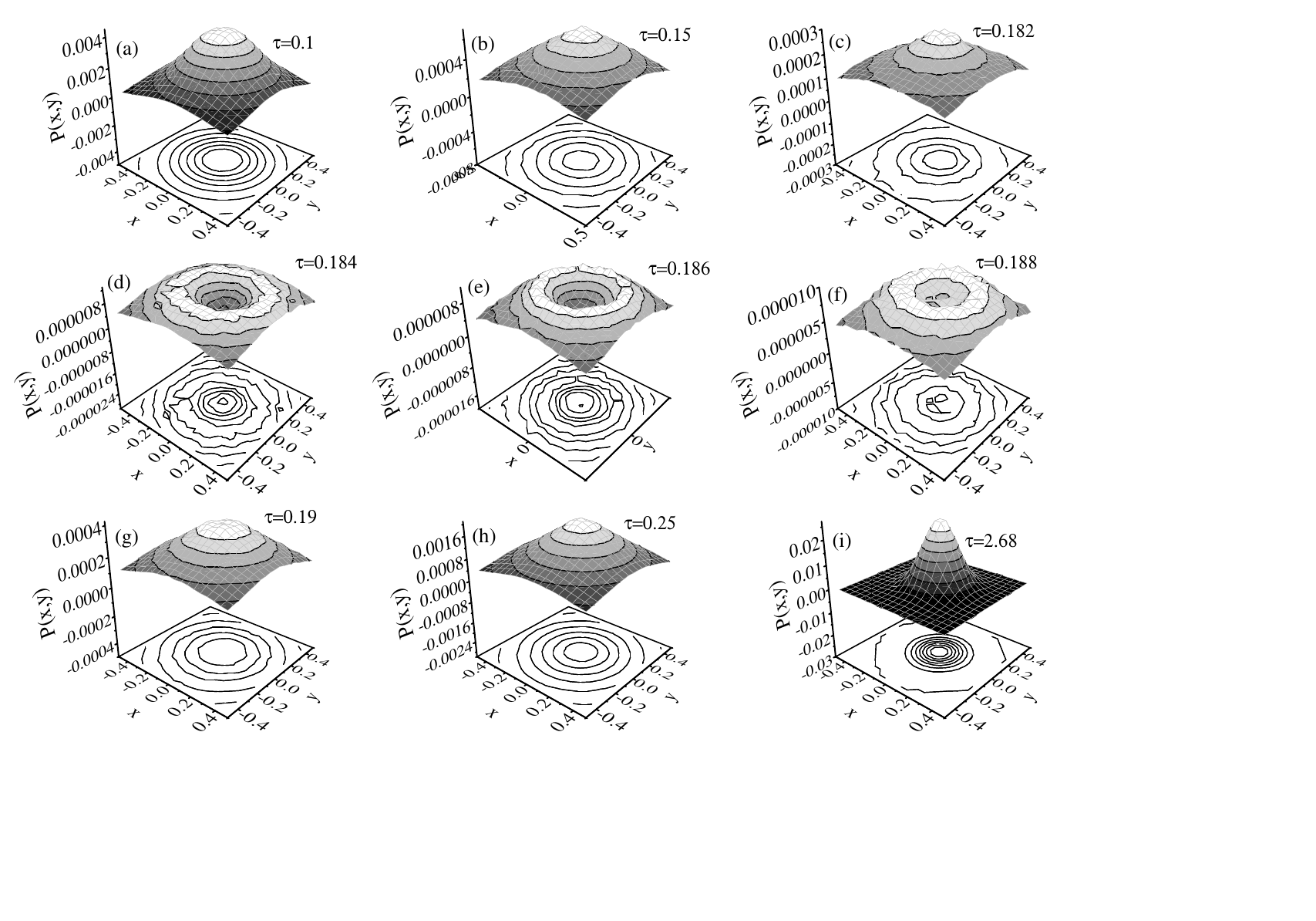}
	\caption{Plot of reduce distribution function ($P(x,y)$ {\it vs} coordinate for the linear system for the parameter set, $\omega^2=2,\Omega_0=5.0$, $\gamma=0.1$, $C=0.42$,  and $k_BT=0.025$.  (Units are arbitrary)}
	\label{fig6}
\end{figure}

\begin{figure}[t!]
	\centering
	\includegraphics[width=1.0\columnwidth,angle=0,clip]{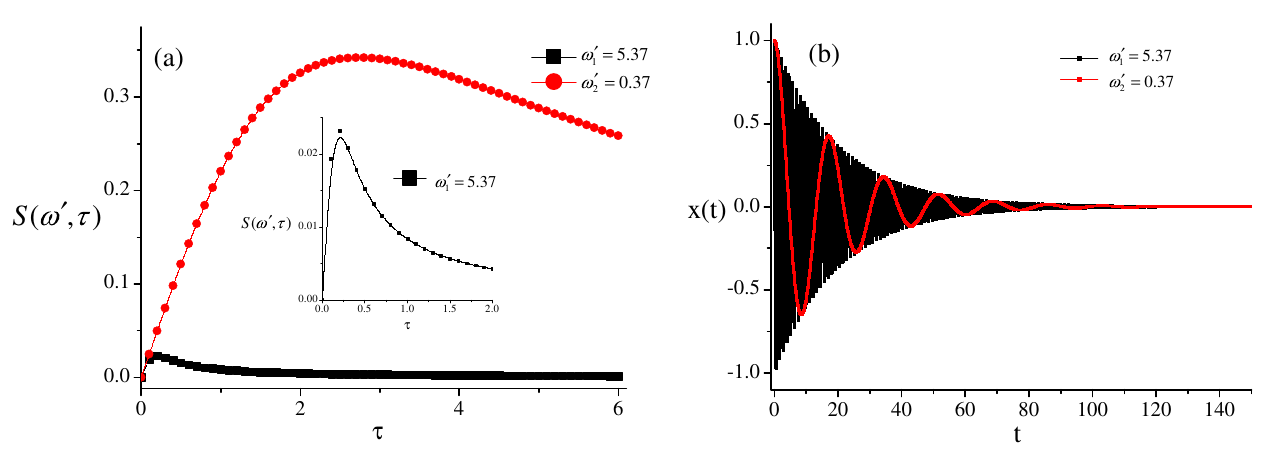}
	\caption{(a) Plot of $S(\omega',\tau)$  {\it vs} $\tau$. (b) Plot coordinate {\it vs} time 
		for equation of motion (\ref{dhs}).	
		(Units are arbitrary)}
	\label{fig7}
\end{figure}

Another point is to be noted here that if the magnetic field is a time independent one ($\eta(t)=0$) then the stationary distribution of particles is a Boltzmann one\cite{baurapre,bag3,bag4} i.e. $\rho(x,y,u_x,u_y)=Ne^{-\frac{E}{k_BT}}$. Here $N$ is the normalization constant and $E(=\frac{mu_x^2+mu_y^2+mu_x^2+m\omega^2(x^2+y^2)}{2})$ corresponds to the total mechanical energy. $E$ is a constant of motion for the equations (\ref{eq33d}) and (\ref{eq34d}) since the magnetic force due to the time independent magnetic field does not work. If this system is coupled with the thermal bath then the stationary state is an equilibrium one with Boltzmann distribution function. In other words, the thermal noise induced transition is not possible in the presence of a time independent Magnetic field.
We now consider the other case. Based on the equation of motion (\ref{eq12}-\ref{eq13}) the rate of change of energy ($E$) of the system  with time in the absence of thermal bath can be read as

\begin{equation}\label{eq23}
	\dot{E} =\frac{\dot{\eta}m}{2}(y\dot{x}-x\dot{y})
\end{equation}
\noindent
Thus the energy of this system is not a constant of motion due to the induced electric field. Therefore on coupling this system with the thermal bath, the stationary state may be a steady state one with the non Boltzmann type distribution function. In other words, the random force due to the fluctuating magnetic field is not related to the damping strength and therefore the stationary state must be a steady state. As a signature of that, the stationary distribution function may depend on the strength of both the time independent and the dependent parts of the magnetic field as implied in Figs. 1 and 4. Thus both the parts are important to appear the noise induced transition where the cyclotron like motion is sustained in the presence of dissipation by virtue of the induced electric field.
\begin{figure}[t!]
	\centering
	\includegraphics[width=1.0\columnwidth,angle=0,clip]{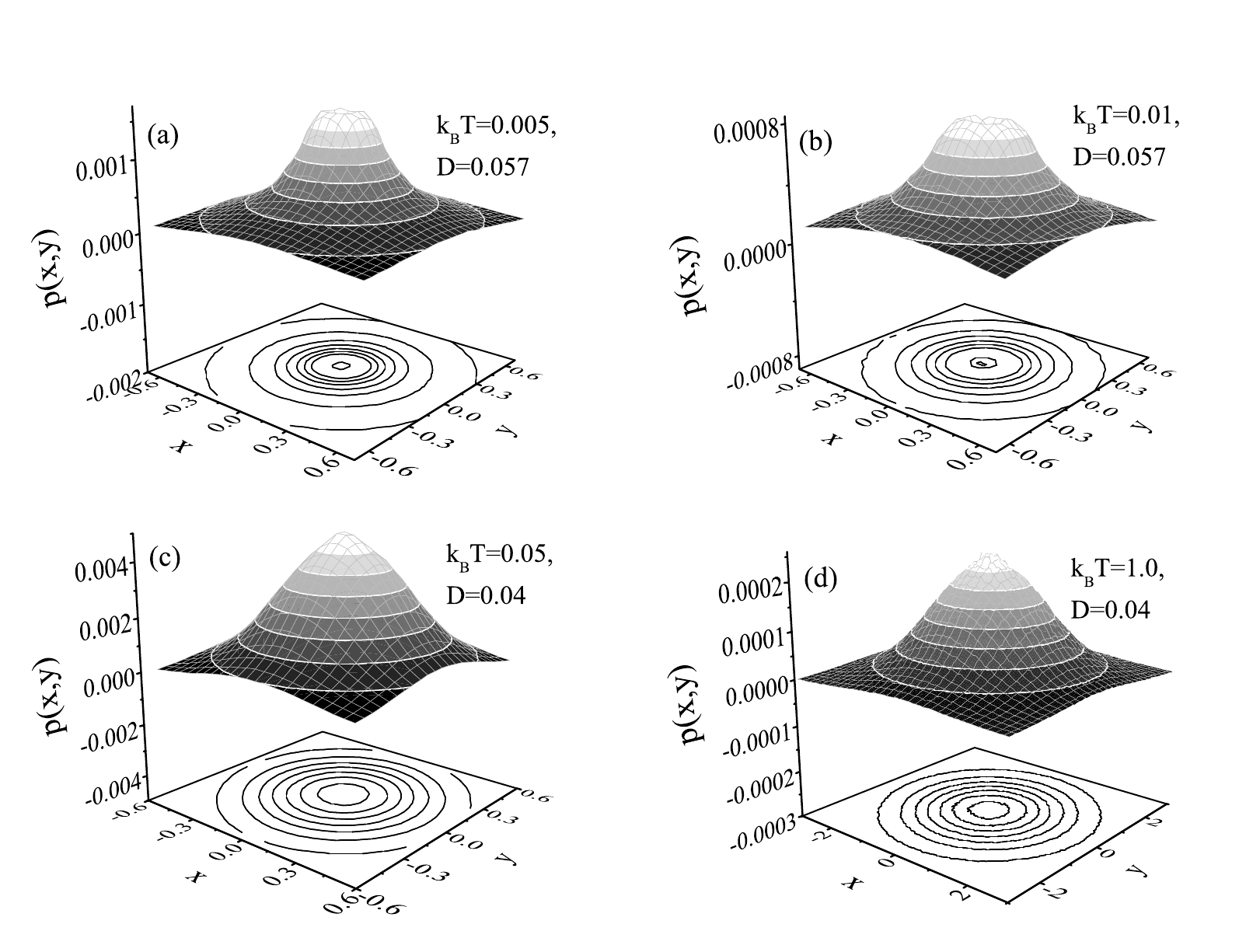}
	\caption{Demonstration of distribution of charged particles in space for a linear system for the parameter set,  $\tau=0.2, \gamma=0.1, \omega^2=2$ and $\Omega_0=7.0$. (Units are arbitrary)}
	\label{fig8}
\end{figure}
\subsection{The damping strength and the noise correlation time induced transition}

Since the stationary state in the presence of a fluctuating magnetic field is a steady state one then the distribution of particles may depend on the damping strength and the correlation time of the fluctuating magnetic field. The fate of the panels (c) and (e) in Fig.~1 at a relatively higher damping strength or correlation time has been demonstrated in Fig.~5. It implies that if the energy dissipation dominates over the work done by the induced electric field then the cyclotron like motion may not sustain at the steady state and the distribution may be close to an equilibrium one. A similar situation may appear even for a low damping strength provided the variance of the fluctuating magnetic field becomes small at a relatively large correlation of the noise. Thus Fig.~5 corroborates the explanation as given in the previous subsection.  

From Figs.~1 and 5 it is apparent that both the damping strength and the noise correlation time can induce
transition. We have checked that it is not possible for the systems like (\ref{ad1}-\ref{ad2}) and (\ref{ad3}), respectively. Then the damping strength and the noise correlation time induced
transition in the fluctuating magnetic field driven harmonic oscillator seems to be a unique signature of the field.

\subsection{The autonomous stochastic resonance induced transition} 

The autonomous stochastic resonance(ASR) may appear for the present system due to the presence of the colored fluctuating magnetic field. In a recent study\cite{bag6} it has been shown that the field may induce resonant activation. Then one may be interested to examine the signature of the ASR on the distribution of particles. In this context, we have calculated the reduced distribution for different noise correlation times keeping fixed the noise variance.  Following two time correlation of $\eta(t)$ we have fixed the variance by changing the noise strength as

\begin{equation}
	D(\tau)=C\tau \; \;
\end{equation}

\noindent
where $C$ is a constant.
It is to be noted here that the above relation is used in general\cite{bag6,shm} to study the Ornstein-Uhlenbeck noise induced autonomous stochastic resonance. We have calculated the distribution for different noise strengths corresponding to this relation and plotted it in Fig.~6. Panels (d), (e), and (f) imply the clear signature of the stochastic resonance. It appears around the noise correlation time, $\tau=0.186$. Following Ref.\cite{shm} one may estimate approximately the critical noise correlation time ($\tau_c$) at which the ASR may be significant. Using Eq.(4) in Ref.\cite{shm} we would get two values of $\tau_c$ corresponding to $\omega_1$ and $\omega_2$ as

\begin{equation}
	\tau_{c1}=\frac{1}{\omega_1} \; \;
\end{equation}

\begin{equation}
	\tau_{c2}=\frac{1}{\omega_2} \; \;
\end{equation}

\noindent
Here we have used $n=1$. For the parameter set corresponding to Fig.~6 we have $\tau_{c1}=0.186$ and
$\tau_{c1}=2.68$. It is apparent in Fig.~6 that for the low frequency mode the autonomous stochastic resonance may not be significant for the given damping strength. To understand this we have calculated
the spectrum $S(\omega',\tau)$ (which is the Fourier transform of the two time correlation of $\eta(t)$) \cite{shm} 
by the following relation

\begin{equation}
	S(\omega',\tau)=\frac{2D(\tau)}{1+{\omega'}^2\tau^2} \; \;
\end{equation}
\noindent
and plotted in panel (a) of Fig.~7. It implies that at $\tau_{c2}=2.68$ the ASR would be more significant compared to $\tau_{c2}=0.186$. Then we one may be interested to know the nature of the damped oscillation of the two vibrational modes. To demonstrate the nature we have used the equations of motion, 

\begin{equation}\label{dhs}
	m\ddot{x}=-{\omega'}^2 x-\gamma \dot{x} \; \;.
\end{equation}

\noindent
The solutions of this for $\omega'=\omega_1=$ and $\omega'=\omega_2$ have been demonstrated in panel (b) of Fig.~7. It is apparent in this panel that the low frequency mode 
almost loses the oscillating behavior which is necessary to exhibit the dynamical resonance. Thus Ref.\cite{shm} finds application to a colored noise driven complex system. 

Before leaving this subsection we would mention the pertinent point. We have checked that the autonomous stochastic resonance induced transition is not possible for the systems like (\ref{ad1}-\ref{ad2}) and (\ref{ad3}), respectively. Thus the autonomous stochastic resonance induced transition for the fluctuating magnetic field driven harmonic oscillator may be a noticeable signature of the field.

\begin{figure}[t!]
	\centering
	\includegraphics[width=1.0\columnwidth,angle=0,clip]{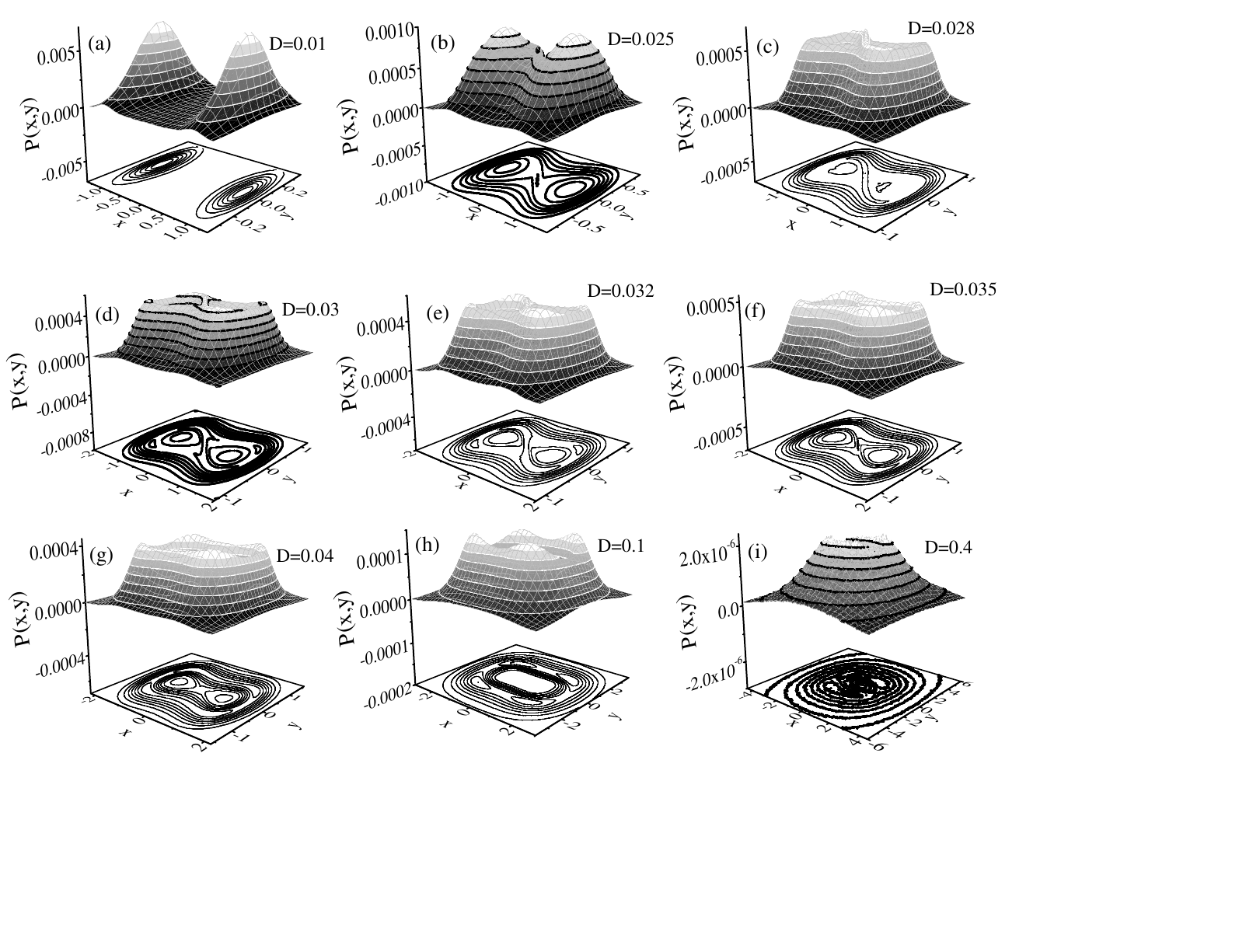}
	\caption{Demonstration of distribution of charged particles in space for the nonlinear system for the parameter set, $a=0.25, b=0.5$, 
		$\tau=0.2, \gamma=0.1, \omega^2=2$, $k_B T=0.025$ and $\Omega_0=7.0$. (Units are arbitrary)}
	\label{fig9}
\end{figure}

\begin{figure}[t!]
	\centering
	\includegraphics[width=1.0\columnwidth,angle=0,clip]{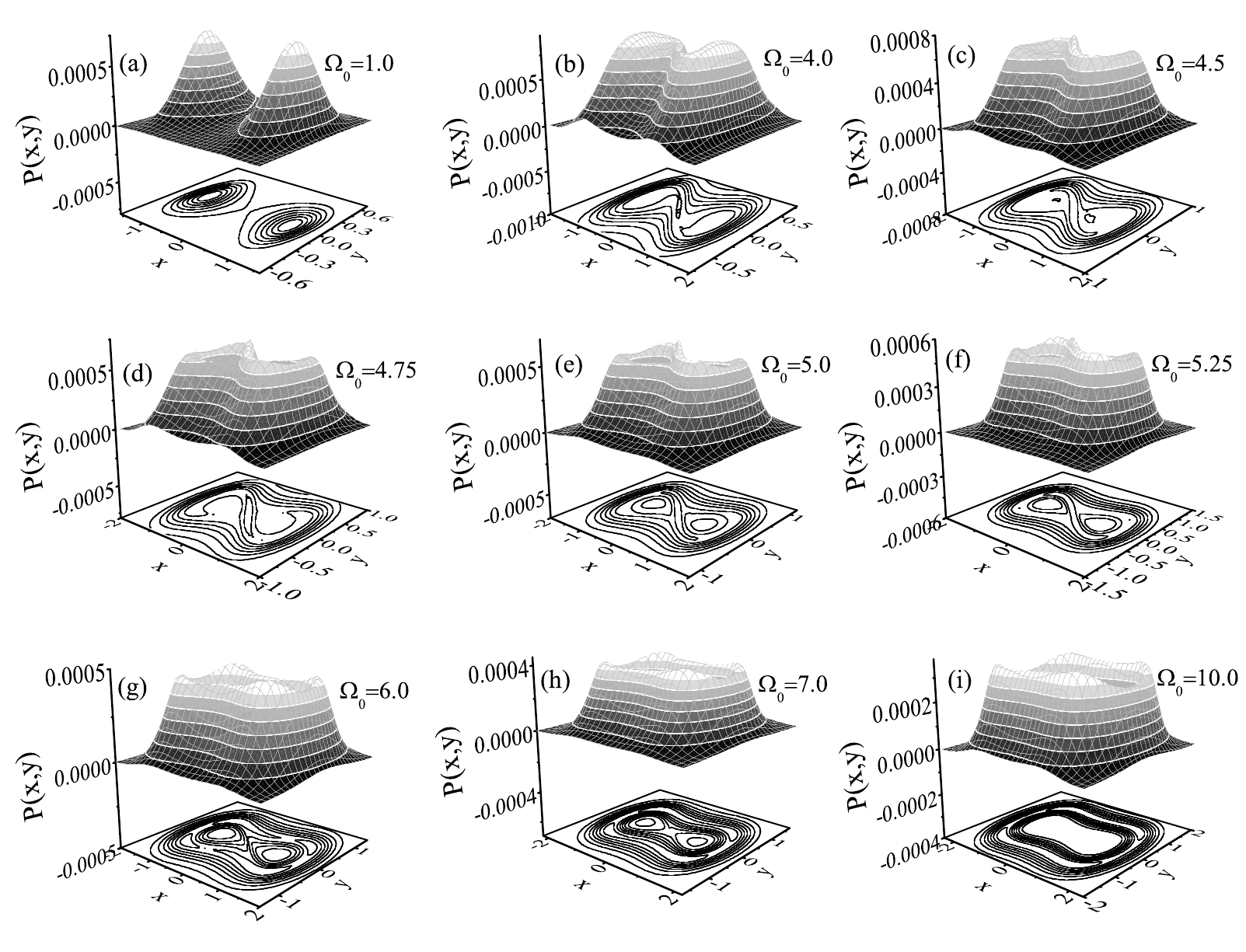}
	\caption{Demonstration of distribution of charged particles in space for the nonlinear system for the parameter set, $a=0.25, b=0.5$,
		$\tau=0.2,  \omega^2=2, D=0.04$, $k_B T=0.025$ and $\gamma=0.1$. (Units are arbitrary)}
	\label{fig10}
\end{figure}

\subsection{Temperature induced transition}

In Fig.~8 we have demonstrated how the distribution of particles may depend on the temperature in the presence of a fluctuating magnetic field. At relatively low temperatures, panels (c) and (e) in Fig.~1 maybe like panels (a) and (b) in Fig.~8, respectively. Then it is apparent that under this circumstance the induced electric field may not be significant to survive cyclotron like motion at steady state and the distribution of particles may be close to Boltzmann type. A similar situation may appear as shown in panels (c) and (d) of Fig.~8 even at relatively high temperatures provided the fluctuating magnetic field is very weak. Thus the magnetic field plays an important role in the noise induced transition in a linear stochastic system where both the deterministic and the stochastic parts are linear functions of the system's
phase space variables. In other words, the temperature can induce transition for the fluctuating magnetic field driven harmonic oscillator. Again it is to be noted here the following point. We have checked that the temperature induced tarnsition is not possible for the systems like (\ref{ad1}-\ref{ad2}) and (\ref{ad3}), respectively. Thus the temperature induced transition for the fluctuating magnetic field driven harmonic oscillator seems to be a special signature of the field.

\begin{figure}[t!]
	\centering
	\includegraphics[width=1.0\columnwidth,angle=0,clip]{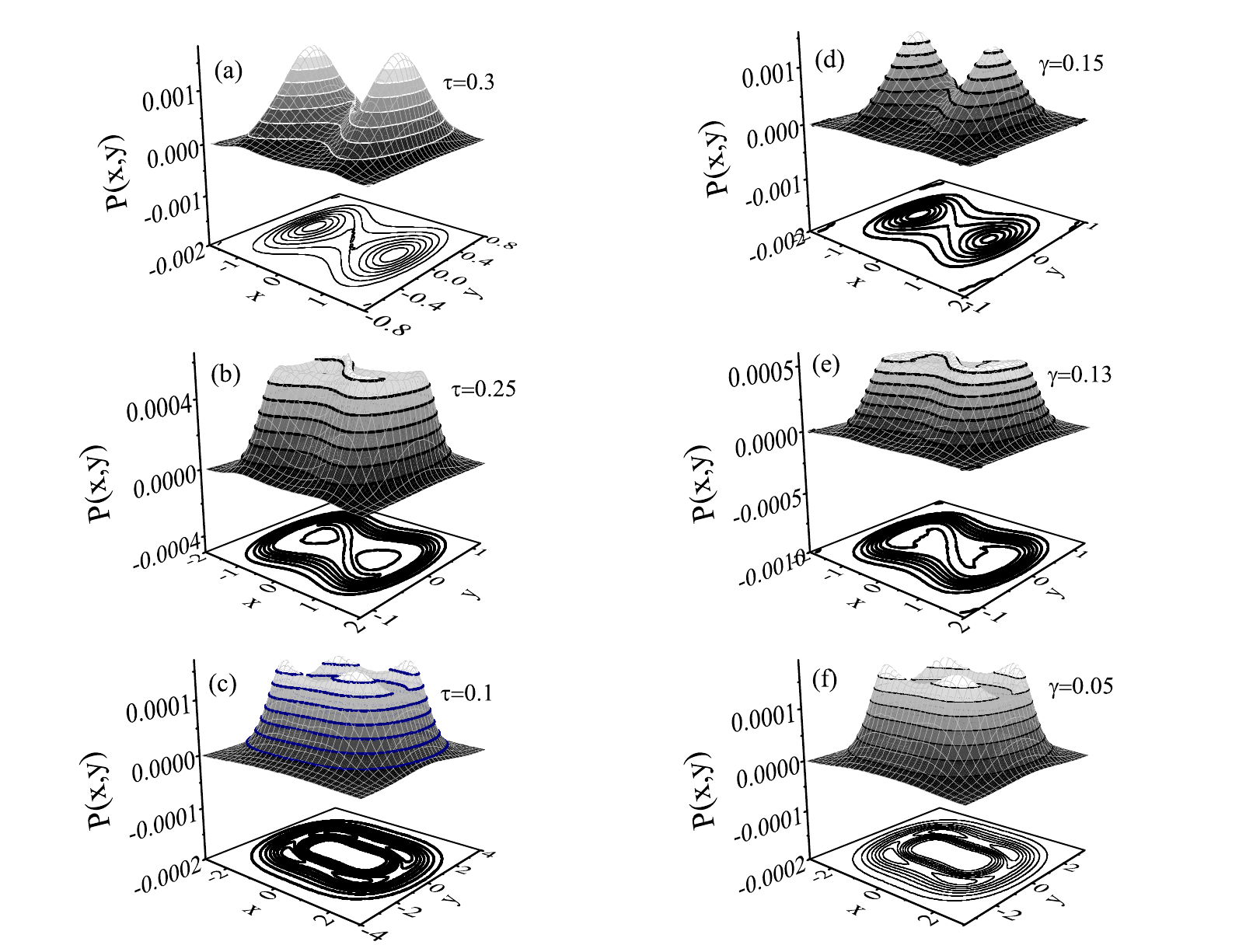}
	\caption{Demonstration of distribution of charged particles in space for the nonlinear system for the parameter set, $a=0.25, b=0.5$,
		$\omega^2=2$, $D=0.04$ and $\Omega=7.0$. (a, b, c) $\gamma=0.1$ (d, e, f) $\tau=0.2$. (Units are arbitrary)}
	\label{fig11}
\end{figure}
\section{Fluctuating magnetic field induced islands formation in a nonlinear system}

The noise induced transition in a linear stochastic system (where both the deterministic and the stochastic parts are linear functions of the system's
phase space variables) strongly motivates us to consider a nonlinear system such as the Brownian motion in two dimensional bi-stable potential energy field. The  potential energy field is given by

\begin{equation}\label{eq11a}
	V(x,y)=a x^4 -b x^2+\omega^2 y^2/2 \; \; \;,
\end{equation}

\noindent
Then the equation for the $x$ component of the motion can be read as

\begin{equation}
	\dot{u_x} =-4 a x^3+2 b x -\gamma u_x +(\Omega_0+\eta(t)) u_y-\frac{\eta y}{2\tau}+\frac{\sqrt{D}y}{2\tau} \zeta(t) + f_x(t)   \; \; \;. \label{eq18}
\end{equation}

\begin{figure}[t!]
	\centering
	\includegraphics[width=1.0\columnwidth,angle=0,clip]{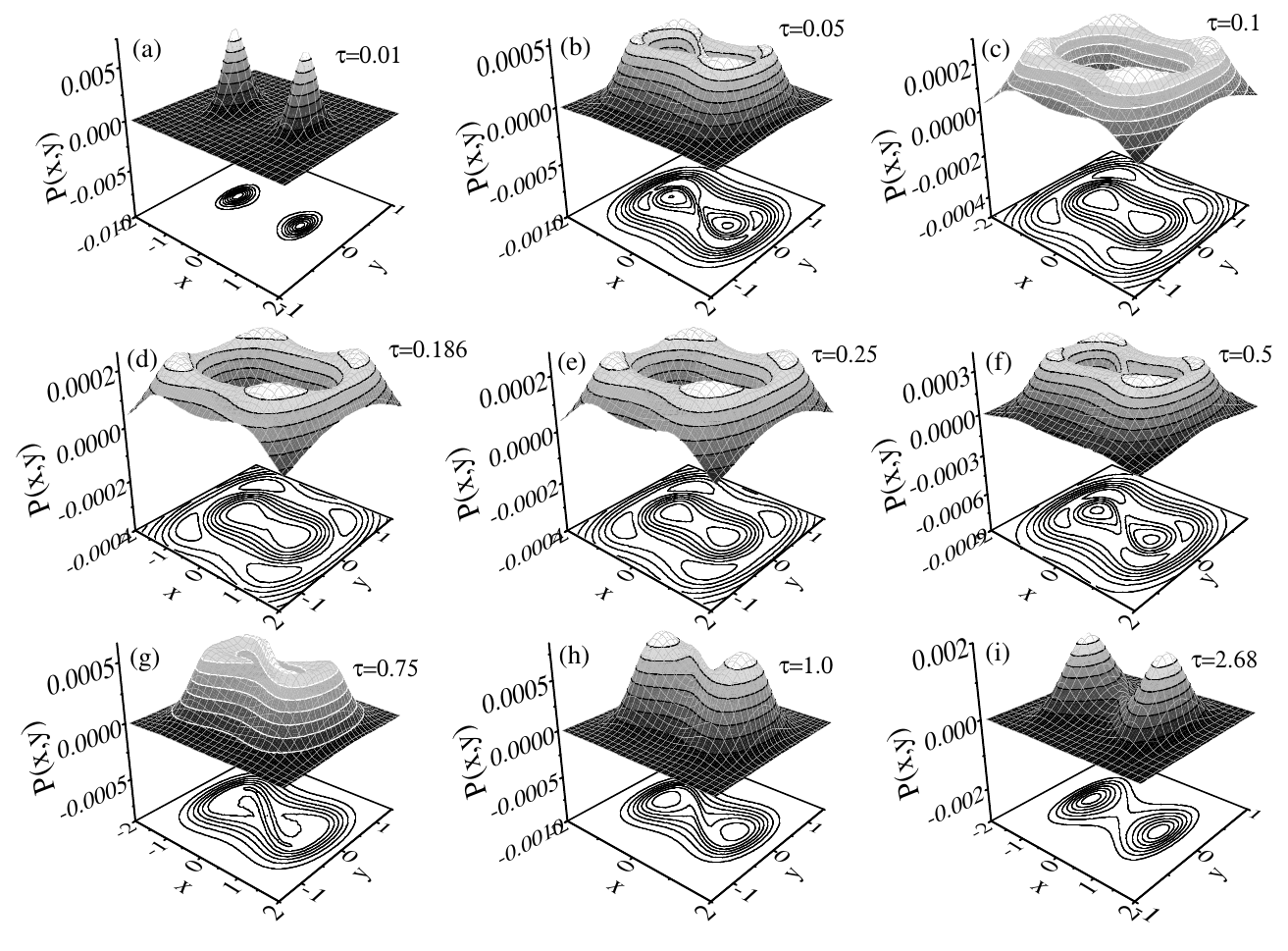}
	\caption{Demonstration of distribution of charged particles in space for the nonlinear system for the parameter set, $a=0.25, b=0.5$, $\omega^2=2$,
		$\gamma=0.1,C=0.43$,$k_B T=0.025$ and $\Omega_0=5.0$. (Units are arbitrary)}
	\label{fig12}
\end{figure}
Solving the above equation along with Eqs.(\ref{eq9}, \ref{eq13}) we have determined the reduced distribution functions, $P(x,y)$ at the stationary state and plotted in Fig.~9. Panel (a) in this figure shows that the distribution of particles in space is closed to the Boltzmann like if the strength of the time dependent magnetic field is relatively very weak. At this condition the system is near to the equilibrium condition $(\eta(t)=0)$.
However, panel (b) is an interesting one which shows that
on further increase in strength of the field, a maximum may appear at the unstable fixed point ($x=0$, $y=0$). It has a similarity with the curves (3, 5, and 6) in panel (a) of Fig.~3. Thus panel 
(a) in Fig.~9 is an example of diffusion controlled noise induced transition. On further increase in strength of the fluctuating field, the distribution of particles may become more complex as shown in other panels of Fig.~9. It is to be noted that one may find a similar kind of distribution of particles (as shown in Fig.~10) on changing the strength of the time independent magnetic field in the presence of fluctuating field at a given strength. From these figures, one may notice the following points. First,
islands may be formed nearby the fixed points. The probability at the stable fixed points may be less compared to the lands as like as the harmonic oscillator.    
Second, the shape of the islands may be different for the nonlinear system. In this context, the formation of S-shape islands around the unstable fixed point (as shown in panel (c) of both figs.9 and 10) seem to be noticeable. Its deviation from the circular shape may be due to an inequivalency between the components of motion even in the absence of a magnetic field. Along the $y$-direction motion is bounded by the harmonic force. But the second derivative of the potential energy with respect to $x$ is negative around the unstable fixed point. It may help to stretch the cyclotron like motion along $x$-direction more compared to another direction. The motion becomes curved at nearby the stable fixed points where both the directions are equivalent in the absence of magnetic field. Thus for the S-shaped cyclotron like motion around the unstable fixed point, the stable fixed points may have an important role. However, the
S-shaped island is confined in 2nd and 4-th quadrants, respectively. This preference may be due to the following fact. In the presence of time dependent magnetic field, the equations of motion may not remain invariant for changing the phase space variables $x,y,u_x,u_y$ and $\eta$ to $-x,-y,-u_x,-u_y$ and $-\eta$. Then it is apparent that this property may offer the preference to the S-shaped island 
in the presence of the significant difference between the components of motion. But as the  cyclotron like motion around the stable fixed points becomes strong (with an increase in contribution from induced electric field) then this preference may reduce as shown by the rest of the panels in Figs.9 and 10. Thus in the presence of a nonlinear potential energy field with unstable fixed point(s), the shape and the number of islands may depend on the interplay among the cyclotron like motions around the stable and unstable fixed points as the signature of the fluctuating magnetic field induced transition with a new mechanism. In other words, an unstable fixed point in a nonlinear system may be an important one to offer the S-shape island (with the preferred orientation ) as well as symmetry reduction.

We now consider the influence of the correlation time of the fluctuating magnetic field on the distribution of charged particles in space. In this context, we demonstrate the distribution function for different $\tau$ in panels (a), (b), and (c) of Fig.~11. It shows that new islands are formed with a decrease in the correlation time. The variance of the fluctuating magnetic field is enhanced for the decrease of $\tau$ and as a result of that, the induced electric field becomes more significant to introduce new islands as explained above. Thus these panels are consistent with Fig.~9. Similarly, the noise induced transition may depend on the damping strength as shown in panels (d), (e), and (f) of Fig.~11. Here it is apparent that the effectiveness of the induced electric field may be more significant (for a given strength of the fluctuating magnetic field) as the energy dissipation becomes weak. Another important point is to be noted here that even for a small change in correlation time of the fluctuating magnetic field or damping strength new islands may be formed. It may be around the unstable fixed point or near the stable fixed point. These observations imply that the change in the structure of the distribution is not due to modification of the potential energy field. In other words, the noise induced transition is due to the interplay between the potential energy field and the cyclotron like motion.

We are now in a position to demonstrate the autonomous stochastic resonance induced transition for the nonlinear system.  Fig.~12 is a relevant one. It is apparent in this figure that the resonance is significant at the critical noise correlation time, $\tau_{c2}=0.186$ which is suggested by the harmonic motion around the bottom of the well as noted in the previous section. Thus the study on the fluctuating magnetic field driven harmonic may not be a mere example as it finds application to understand a nonlinear system which may mimic a relevant experimental situation.  

\section{Conclusion}

We have studied the distribution of particles in space in the presence of a fluctuating magnetic field. The field introduces an unusual type of multiplicative noise terms which are linear functions of coordinate and velocity, respectively. The dynamics with these terms offer a versatile structure  of the distribution function. Our investigation includes the following major points.\\

\noindent
(i) In the presence of a harmonic force field, the distribution function may be folded around the origin with either a maximum or minimum at the origin at a relatively high strength of both time dependent and independent fields. These sharp contrast behaviors of a linear stochastic system (where both the deterministic and the stochastic parts are linear functions of system's
phase space variables) are due to the noticeable signature of the fluctuating magnetic field.

\noindent
(ii) If the potential energy field is a  nonlinear one then new islands like regions are formed in the distribution of charged particles in space with an increase in the strength of the fluctuating magnetic field. The land may be formed near the stable fixed points instead of folding the distribution function in space around the stable fixed points as happens in the linear system. 

\noindent
(iii) The appearance of islands at nearby fixed points seems not to be obvious from the given potential energy field. Even the shape and the number of islands depend on the noise strength. These are also not obvious from the equations of motion. The formation of islands may offer an explanation to describe the phenomenon, the reduction of the current in a semiconductor in the presence of a time dependent magnetic field.

\noindent
(iv) An unstable fixed point in a nonlinear system may be an important one to offer the S-shape island (with the preferred orientation ) as well as symmetry reduction. 

\noindent
(v) The damping strength and the noise correlation time induced
transition in the fluctuating magnetic field driven harmonic oscillator seems to be a unique signature of the field. Similarly, the field driven autonomous stochastic resonance induced transition may be a special one for a linear system where both the deterministic and the stochastic parts are linear functions of the system's phase space variables.

\noindent
(vi) The temperature may induce transition for the fluctuating magnetic field driven harmonic oscillator.

\noindent
(Vii)Finally, based on the present study one may anticipate that the noise induced tarnsition may be possible for an additive noise driven harmonic oscillator in the presence of a time independent magnetic field. But we have checked that the distribution of particle is a uni-modal one even at reletively large strength of the noise. In other words, at the steady state, the cyclotron like motion may not be sustanable against the dissipative force and the noise induced tarnsition is not possible for this system. Then our preliminary observation with a multiplicative noise (which is a linear function of system's coordinate) instead of the additive noise implies that the noise induced transition may be possible for this case. The detail investigation regarding this issue may appear shortly elsewhere.

Before leaving this section we would mention that the present study may be helpful to understand as well as control the conductivity of electrolytic materials.
In the recent past, the study on ion conducting electrolytic materials is a very important area in physics and chemistry being driven by an ever-increasing demand for portable electronic devices. The materials have potential applications in a diverse range of all-solid-state devices, such as rechargeable lithium batteries, flexible electrochromic displays, and smart windows \cite{scros, scros1, scros1I, scros2}.
The properties of the electrolytes are tuned by varying chemical composition to a large extent and hence are adapted to the specific
needs \cite{angel,angel1}. High ionic conductivity is needed for optimizing the glassy electrolytes in various applications. It would be very interesting if one can tune the ionic conductivity according to the specific need by a physical method. One may investigate the issue in the presence of the Lorentz force. Although a time independent magnetic field can not activate the particle to cross the barrier it may modulate the frequency of the dynamics. It is to be noted here that a time dependent magnetic field may introduce an induced electric field to activate the particle. Another way is the direct application of an electric field which may be helpful in the case when a very high rate of barrier crossing is necessary. In the very recent Refs.~\cite{katsuki,pere, pereI,pereII,telang,vdo,bag3,bag4,bag5,bag6,physa} it has been shown that the conductivity of an electrolytic material can be tuned by an applied magnetic field. Electron tunneling in quantum wire in the presence of a magnetic field has been studied in Ref.~\cite{katsuki}.  Tunneling ionization of impurity centers in semiconductors was investigated \cite{pere, pereI,pereII} in the presence of a magnetic field. The effect of a magnetic field on the electron transport of GaAs quantum wire was studied very recently \cite{telang} in the presence of an electric field. Based on the magneto-tunneling spectroscopy a noninvasive and nondestructive probe has been used to produce two-dimensional spatial images of the probability density of an electron confined in a self-assembled semiconductor quantum dot \cite{vdo}. The technique exploits the effect of the classical Lorentz force on the motion of a tunneling electron. In this experiment, extremum behavior is observed in the variation of the tunneling current with the magnetic field strength. A similar kind of behavior also has been reported in Refs. \cite{bag3,bag5} in the variation of the barrier crossing rate constant as a function of the strength of the time independent magnetic field (TIMF). This optimum behavior has been observed at a low damping regime \cite{bag3,bag5}. However, in Refs. \cite{bag4, physa} it has been shown that one may tune the barrier crossing rate by virtue of fluctuating magnetic field. It is to be noted here that the current in the semiconductor may become small for a given amplitude of the oscillating magnetic field. This observation has been explained based on the phenomenon, dynamical localization. This phenomenon has been of growing interest \cite{dloc, dlocI,dlocII, dlocIII, dlocIV} because of its very relevance in the response of the electron transport in mesoscopic systems to external fields \cite{dloc1, dloc1I, dloc1II}. In this context, the present study may be a very relevant one. Based on it one can account for the reduction of the current in the semiconductor. The formation of islands may reduce the conductivity of electrolytes in the presence of a time dependent magnetic field.

\appendix
\section{Decoupling of equations (\ref{eq12}-\ref{eq13}) of motion}

One may formally decouple Eqs.(\ref{eq12}-\ref{eq13}) considering homogenous and particular solutions of the respective equations.
General solution of these equations can be read as

\begin{equation}
	x=e^{-Kt}\alpha(t)+x_{p}.\label{a1}
\end{equation}

\begin{equation}
	y=e^{-Kt}\alpha(t)+y_{p}.\label{a2}
\end{equation} 

\noindent
where $K=\frac{\gamma_0}{2}$ and  $e^{-Kt}\alpha(t)$ is the solution of the equations at the limit, $\eta(t)=0, f_x(t)=0, f_y(t)=0, D=0$ and $\Omega_0=0$. $\alpha(t)$ has three forms\cite{bag6} for the conditions (i) $\omega>K$, (ii) $\omega<K$ and (iii) $\omega=K$, respectively. Then the respective particular solutions $x_p$ and $y_p$ are defined as 

\begin{equation}
	Lx_p=-\left(\Omega_{0}+\eta(t)\right)u_{y}+\dfrac{\dot{\eta}y}{2}+f_{x}(t)\label{a3}
\end{equation}

\begin{equation}
	Ly_p=-\left(\Omega_{0}+\eta(t)\right)u_{x}-\dfrac{\dot{\eta}x}{2}+f_{y}(t)\label{a4}
\end{equation} 

\begin{equation}
	x_p=L^{-1}\left[-\left(\Omega_{0}+\eta(t)\right)u_{y}+\dfrac{\dot{\eta}y}{2}+f_{x}(t)\right] \label{a5}
\end{equation} 

\begin{equation}
	y_{p}=L^{-1}\left[-\left(\Omega_{0}+\eta(t)\right)u_{x}-\dfrac{\dot{\eta}x}{2}+f_{y}(t)\right]\label{a6}
\end{equation}

\noindent 
where 
\begin{equation}
	L=\dfrac{d^{2}}{dt^{2}}+\gamma_{0}\dfrac{d}{dt}+\omega^{2} \; \;. \label{a6a}
\end{equation}

\noindent
Then using Eq.(\ref{a2}) with the defination of $y_p$ in Eq.(\ref{eq12}) 
for $u_y=\dfrac{dy}{dt}$ we have

\begin{equation}
	\begin{array}{rcl}
		\dot{u}_{x} & = & -\omega^{2}x-\gamma_{0}u_{x}+\left(\Omega_{0}+\eta(t)\right)\left[\dfrac{d}{dt}
		\left(e^{-Kt}\alpha_{1}(t)\right)+\dfrac{d}{dt}\left(L^{-1}\left\{ -\left(\Omega_{0}+\eta(t)\right)u_{x}-\dfrac{\dot{\eta}x}{2}+f_{y}(t)\right\} \right)\right]\\
		&  & +\dfrac{\dot{\eta}}{2}\left[e^{-Kt}\alpha_{1}(t)+L^{-1}\left\{ -\left(\Omega_{0}+\eta(t)\right    )u_{x}-\dfrac{\dot{\eta}x}{2}+f_{y}(t)\right\} \right]+f_{x}(t)
	\end{array} \; \; \;.  
	\label{a7}
\end{equation}

\noindent
Similarly one may read that

\begin{equation}
	\begin{array}{rcl}
		\dot{u}_{y} & = & -\omega^{2}y-\gamma_{0}u_{y}-\left(\Omega_{0}+\eta(t)\right)\left[\dfrac{d}{dt}
		\left(e^{-Kt}\alpha_{1}(t)\right)+\dfrac{d}{dt}\left(L^{-1}\left\{ \left(\Omega_{0}+\eta(t)\right)u_{y} +\dfrac{\dot{\eta}y}{2}+f_{x}(t)\right\} \right)\right]\\
		&  & -\dfrac{\dot{\eta}}{2}\left[e^{-Kt}\alpha_{1}(t)+L^{-1}\left\{ \left(\Omega_{0}+\eta(t)\right)    u_{y}+\dfrac{\dot{\eta}y}{2}+f_{x}(t)\right\} \right]+f_{y}(t)
	\end{array} \; \; \;.
	\label{a8}
\end{equation}

To rearrange these equations one may look into the operator, $L$. 
Then Eq.(\ref{a6a}) can be rearranged as

\begin{equation}
	L=1+M  \label{a6b}
\end{equation}

\noindent
with

\begin{equation}
	M=\dfrac{d^{2}}{dt^{2}}+\gamma_{0}\dfrac{d}{dt}+\omega^{2}-1 \label{a6c}
\end{equation}

\noindent
Thus $L$, as well as $M$, are linear operators. The operator, $L^{-1}$ can be expressed as 

\begin{equation}
	L^{-1}  = \sum_{k=0}^{\infty}(-1)^kM^k \; \; \; \label{a6d}
\end{equation}

\noindent
The above equation implies that $L^{-1}$ is also a linear operator. 
Then using this property in Eqs.(\ref{a7}-\ref{a8}) we have 

\begin{equation}
\begin{array}{rcl}
\dot{u}_{x} & = & -\left[\omega^{2}+\left(\Omega_{0}+\eta(t)\right)\dfrac{d}{dt}\left(L^{-1}\dfrac{\dot{\eta}}{2}\right)+\dfrac{\dot{\eta}}{2}L^{-1}\left(\dfrac{\dot{\eta}}{2}\right)\right]x\\
&  & -\left[\gamma_{0}+\left(\Omega_{0}+\eta(t)\right)\dfrac{d}{dt}\left(L^{-1}\left\{ \Omega_{0}+\eta(t)\right\} \right)+\dfrac{\dot{\eta}}{2}L^{-1}\left\{ \Omega_{0}+\eta(t)\right\} \right]u_{x}\\
&  & +\left(\Omega_{0}+\eta(t)\right)\left[\dfrac{d}{dt}\left(e^{-Kt}\alpha_{1}(t)\right)+\dfrac{d}{dt}\left(L^{-1}\left\{ f_{y}\left(t\right)\right\} \right)\right]+\dfrac{\dot{\eta}}{2}\left[e^{-Kt    }\alpha_{1}(t)+L^{-1}\left\{ f_{y}\left(t\right)\right\} \right]+f_{x}(t)
\end{array}\label{a9}
\end{equation}
\noindent
and 

\begin{equation}
\begin{array}{rcl}
\dot{u}_{y} & = & -\left[\omega^{2}+\left(\Omega_{0}+\eta(t)\right)\dfrac{d}{dt}\left(L^{-1}
\dfrac{\dot{\eta}}{2}\right)+\dfrac{\dot{\eta}}{2}L^{-1}\left(\dfrac{\dot{\eta}}{2}\right)\right]y\\
&  & -\left[\gamma_{0}+\left(\Omega_{0}+\eta(t)\right)\dfrac{d}{dt}\left(L^{-1}\left\{ \Omega_{0}+    \eta(t)\right\} \right)+\dfrac{\dot{\eta}}{2}L^{-1}\left\{ \Omega_{0}+\eta(t)\right\} \right]u_{y}\\
&  & -\left(\Omega_{0}+\eta(t)\right)\left[\dfrac{d}{dt}\left(e^{-Kt}\alpha_{1}(t)\right)+\dfrac{d}    {dt}\left(L^{-1}\left\{ f_{x}(t)\right\} \right)\right]-\dfrac{\dot{\eta}}{2}\left[e^{-Kt}\alpha_{1}    (t)+L^{-1}\left\{ f_{x}\left(t\right)\right\} \right]+f_{y}(t) \; \;.
	\end{array}\label{a10}
\end{equation}

\vspace{2 cm}

\noindent
\textbf{Acknowledgment}

\noindent
L.R.R.B. acknowledges financial
support from DST-INSPIRE from the Department of Science
and Technology, Government of India. M.B. is thankful to CSIR, Government of India for financial
support.

\bibliographystyle{elsarticle-num}
\bibliography{references}
\end{document}